\documentclass[smallextended]{svjour3}  
\usepackage{subcaption}
\usepackage[labelformat=parens,labelsep=quad, skip=3pt]{caption}
\usepackage{graphicx}
\usepackage{amssymb}
\usepackage[utf8]{inputenc}
\usepackage{array,url}
\usepackage{tabularx}
\usepackage[table,xcdraw]{xcolor}
\usepackage{multirow}
\usepackage{multicol,url,color}
\usepackage{rotating,booktabs}
\usepackage{adjustbox}
\usepackage{amsmath,amssymb,amsfonts}
\usepackage{algorithm}
\usepackage[noend]{algpseudocode}
\usepackage[multiple]{footmisc}
\usepackage{bbding}
\usepackage{comment}
\usepackage{textcomp}
\usepackage{natbib}
\usepackage{pifont}
\makeatother
\usepackage{float}
\usepackage{fancyhdr}
\usepackage{mathtools} 
\usepackage{graphics}
\usepackage{placeins}
\usepackage{enumitem}
\usepackage{color}
\usepackage[bookmarks=false]{hyperref}
\usepackage{colortbl}

\usepackage{hhline}
\title{An Analytical Survey on Recent Trends in High Dimensional Data Visualization}
\author{Alexander Kiefer \and Md. Khaledur Rahman}

\institute{A. Kiefer, Indiana University Bloomington, \email{alkiefer@iu.edu} 
           \and
           M. K. Rahman, 
           Indiana University Bloomington, \email{morahma@iu.edu}
}

\usepackage{geometry}
\usepackage{graphicx}

\begin{document}

\maketitle

\begin{abstract}
Data visualization is the process by which data of any size or dimensionality is processed to produce an understandable set of data in a lower dimensionality, allowing it to be manipulated and understood more easily by people. The goal of our paper is to survey the performance of current high-dimensional data visualization techniques and quantify their strengths and weaknesses through relevant quantitative measures, including runtime, memory usage, clustering quality, separation quality, global structure preservation, and local structure preservation. To perform the analysis, we select a subset of state-of-the-art methods. Our work shows how the selected algorithms produce embeddings with unique qualities that lend themselves towards certain tasks, and how each of these algorithms are constrained by compute resources.
\end{abstract}


\section{Introduction}
\quad According to the International Data Corporation (IDC), a leading market analytics company within the information technology industry, it is projected that ``the Global Datasphere will grow from 33 Zettabytes (ZB) in 2018 to 175 ZB by 2025"~\citep{rydning2018digitization}. With this significant increase in global information production come a number of opportunities for businesses to make more informed decisions using mathematics, statistics, and machine learning. However, with this increase in data, also come a number of challenges that must be addressed in order to reap its possible benefits, chief among those being the difficulty associated with the interpretation and understanding of raw data produced by technology. Data visualizations are capable of providing a great depth of information with high density, making them crucial tools in conveying the underlying meaning of data to people.

Data visualization is the process by which data of any size or dimensionality is processed to produce an understandable set of data in a lower dimensionality, allowing it to be manipulated and understood more easily by people. The typical dimension for visualization is 2D or 3D in picture/image format. Beyond 2D or 3D, the visualization is not easily perceivable by human beings. Good quality visualizations allow us to extract a wealth of meaningful information at first hand, as the famous saying goes``a picture is worth a thousand words"~\citep{pinsky2000picture}.

There are a handful number of dimensionality reduction techniques that help us reduce the high dimensional data to 2D or 3D data for visualization. The dimensionality reduction algorithms which we will focus on typically fall into two groups, those which attempt to maintain the global structure of the data and those which maintain local distances over global distance~\citep{mcinnes2018umap}. In this paper, of the algorithms discussed, TriMap~\citep{amid2019trimap} falls into the first category, while UMAP~\citep{mcinnes2018umap}, LargeVis~\citep{tang2016visualizing}, and t-SNE~\citep{maaten2008visualizing} fall into the second. Other research fields, such as computational biology and graph embedding, are being significantly benefited from the visualization of high-dimensional data by using these methods~\citep{becht2019dimensionality,kobak2019art,tsitsulin2018verse,rahman2020force2vec}. The focus of our study will be on the effectiveness and applications of each of these algorithms for high-dimensional data visualization.

The dimensionality reduction techniques mainly work on the high-dimensional numeric data. However, if we have graphs/networks, we cannot apply them directly though they are also considered as high-dimensional structured datasets~\citep{erdos1965dimension}. Thus, in close association with the previous process, graph visualization is a method by which data of low dimensionality (2D or 3D) is used to generate visual interpretations of graphs or networks, making the relationships within the data more easily understood by users. These types of datasets are structured and we can extract valuable information from the visualizations. In this survey, we focus on analyzing the runtimes, memory consumption and the quality of layouts for some general purpose graph visualization techniques. Force-directed techniques are widely used for graph visualization. Thus, we will primarily focus on some state-of-the-art force-directed methods for large-scale graph visualization along with a dimensionality reduction technique for graph visualization.

The difference between the high-dimensional data point and graph visualization is that we only show the positions of the data points for the former whereas edges of a graph are also drawn in 2D/3D space for the latter. Though there are multiple survey papers in the literature for data visualization, an analytical study showing the performance comparison of existing methods is missing~\citep{liu2014survey,chen2019survey,van2009dimensionality}. In this survey paper, we fill this gap by analyzing runtimes, memory consumption, and aesthetic quality scores of some representative visualization methods from the two categories. To make a fair comparison, we run all the methods in the same server machine, and carefully analyze the results. We summarize the main focus of our contributions as follows:

\begin{itemize}
    \item We select a set of representative state-of-the-art visualization methods based on the popularity, and present the underlying methods in a concise but understandable format. Our survey covers two fundamental visualization areas: (i) vector data, and (ii) graphs or networks, which provide a global picture of the respective visualization field.
    
    \item We compare all the methods in terms of runtimes, memory consumption, and quality scores to analyze the advantage and disadvantage of using a respective method. 
    
    \item Finally, we analyze the results to provide recommendations to users or researchers so that they can get an idea regarding which method to use for a given set of resources.
\end{itemize}

\section{Preliminaries}

\quad We represent high-dimensional vector data by  $X\in \mathbf{R}^ {n\times d}$, where $n$ is the number of entries and $d$ is the dimension of each data entry. Hence, $x_i$ represents $d$-dimensional vector of $i$-th entry of $X$. We represent the lower dimensional projection of $X$ by $Y\in \mathbf{R}^ {n\times 2}$, where $y_i$ represents $(x, y)$ coordinates of $i$-th entry in 2D space. Let $G(V, E)$ be a graph, where $V$ is the set of vertices and $E$ is the set of edges. Unless otherwise mentioned, we use the same definition of $Y$ to represent the layout of graphs in 2D space.

Asymmetric high dimensional data points belong to vector data category when they have no explicit structural connection such as images, words, etc. On the other hand, data points having structural information belong to Network category. Without loss of generality, graph and network bear the same meaning in our study. Thus, we use these two terms interchangeably. 

\begin{figure}
    \centering
    \includegraphics[width=0.60\linewidth]{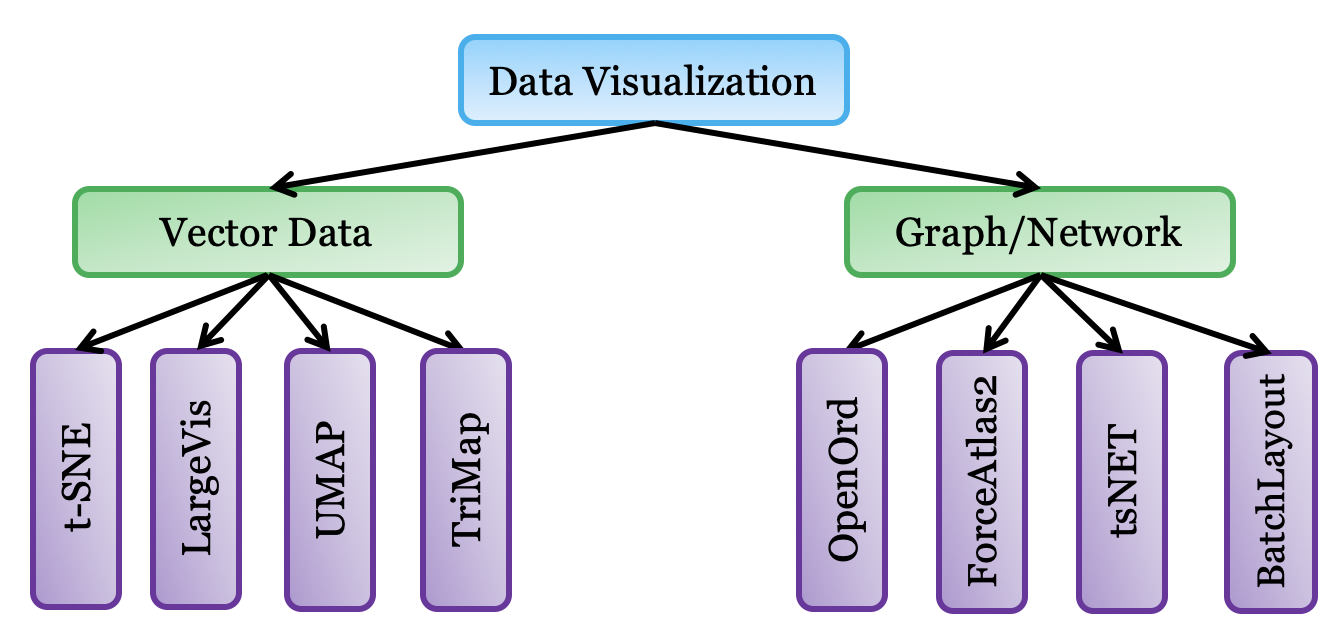}
    \caption{Data visualization hierarchy}
    \label{fig:datavishierarchy}
\end{figure}

\section{High Dimensional Data Visualization}
Data can come from a variety of sources which can be structured or unstructured. Different methods have been proposed in the literature to visualize both types of data. In this survey, we discuss visualization methods focusing on two branches: (i) vector data, and (ii) graph data. We select a few methods to analyze both types of visualization techniques. Fig. \ref{fig:datavishierarchy} shows a hierarchy of our selected data visualization techniques.

\subsection{Vector Data Visualization} 
\quad Generally, dimensionality reduction techniques are used to project high dimensional data points to lower dimensions. This lower dimensional projection captures the intrinsic properties of the vector data as much as possible for readable visualization. Similar data points are clustered together whereas dissimilar points are placed apart. We discuss some of the state-of-the-art methods as follows.

\subsubsection{t-SNE \citep{maaten2008visualizing}}
\quad For high-dimensional vector data visualization, t-SNE is a pioneering tool and a progenitor of most modern dimensionality reduction tools. The algorithm significantly improved upon ones prior to it by allowing for much higher dimensionality in data sets and, overall, better visualization by reducing the tendency of points to cluster at the center of the graph, which is often referred to as the crowding-out problem. By employing a variation of Stochastic Neighbor Embedding \citep{NIPS2002_2276}, t-SNE is significantly easier to optimize, as compared to prior methods, increasing both the efficiency and quality of graphs, with a computational complexity modeled by $O(N^{2})$.

To give a very general overview of t-SNE, high-dimensional data sets are converted into a pairwise similarities matrix to allow for construction of accurate and meaningful visualizations. Stochastic neighbor embedding begins by converting the high-dimensional Euclidean distances within the input data into a set of conditional probabilities, which represents the similarity between the data points. The conditional probability, $p_{j|i}$, can be modeled by Eqn. \ref{eqn:label1} as follows.

\begin{equation}
\label{eqn:label1}
p_{j|i}=\frac{exp(-{||x_{i}-x_{j}||}^{2}/2{\sigma}_{i}^{2})}
{\sum\nolimits_ {k \neq i} exp(-{||x_{k}-x_{l}||}^{2}/2{\sigma}_{i}^{2})}
\end{equation}

Where the similarity between two data points $x_{i}$ and $x_{j}$ is equal to the conditional probability that $x_{i}$ is the neighbor of $x_{j}$, given that neighbors are picked under a Gaussian centered at $x_{i}$ in proportion to their probability density. ${\sigma}_{i}$ is the variance of the Gaussian on $x_{i}$. The values of $p_{i|i}$ is set to zero in order to constrain the model to pairwise similarities.

Next, to find the low-dimensional results, $y_{i}$  and $y_{j }$ of $x_{i}$ and $x_{j}$ another conditional probability, $q_{j|i}$ is calculated. However, in this instance, the variance of the Gaussian is set to $\frac{1}{\sqrt{2}}$. Thus, the similarity between the mapped points $y_{i}$  and $y_{j}$ is modeled by
\newline
\begin{equation}
\label{eqn:label2}
q_{j|i}=\frac{exp(-{||y_{i}-y_{j}||}^{2})}
{\sum\nolimits_ {k \neq i}
exp(-{||y_{k}-y_{l}||}^{2})}
\end{equation}

If the values of $q_{j|i}$ and $p_{j|i}$ are equal, then the mapped points $y_{i}$  and $y_{j}$ correctly model the original data points $x_{i}$ and $x_{j}$. However, to improve efficiency, SNE allows for a small amount of variance between $q_{j|i}$ and $p_{j|i}$, which it optimizes by minimizing a Kullback-Leibler divergence between a joint probability distribution, $P$, in high-dimensional space and a joint probability distribution, $Q$, in low-dimensional space. The cost function of this method is given by the model
\newline
\begin{equation}
\label{eqn:label3}
C= KL(P||Q)=\sum\limits_{i} \sum\limits_{j} p_{j|i}log \frac{p_{j|i}}{q_{j|i}}
\end{equation}

Where $P_{i}$ is the probability distribution, given data point $x_{i}$, over all other data points and $Q_{i}$ is the same, but given data point $y_{i}$. Due to the unsymmetrical nature of the Kullback-Leibler divergence, if the mapped points used to represent nearby data-points are far apart, they will have a higher cost function associated with them. However, if the mapped points used to represent far apart data-points are close together, they will have a lower cost function associated with them.

The final parameter that must be set is the variance $\sigma_{i}$ of the Gaussian. It is beneficial to pick a smaller value for $\sigma_{i}$ when the region is more dense, and a larger one when it is more sparse. The value $\sigma_{i}$ is found through a binary search with perplexity $P_{i}$ defined by the user. One of the superior qualities of symmetric SNE, as compared to asymmetric SNE, is that the gradient function \citep{pmlr-v2-cook07a} is much simpler, decreasing computation time.
\newline
\begin{equation}
\label{eqn:label4}
\frac{\delta C}{\delta y_{i}}=4 \sum\limits_{j} (p_{ij}-q_{ij})(y_{i}-y_{j})
\end{equation}

\subsubsection{LargeVis \citep{tang2016visualizing}}
\quad Developed in 2016, LargeVis aims to solve the problem of the high computational cost associated with dimensionality reduction problem such as t-SNE. To give a very general overview of the LargeVis method, first, an accurately approximated $k$-nearest neighbor graph is generated from the input data. After this, the graph is then projected in a low dimensional space in a layout which is easily readable by the viewers. By using a principle probabilistic model, the dimensional reduction of the data can be optimized using an asynchronous stochastic gradient descent with a linear time complexity. Some advantages that LargeVis has over other methods are higher stability of its hyper-parameters over a large variety of datasets and greater efficiency and effectiveness as compared to t-SNE, with an overall computational complexity of $O(dsN)$, where $d$ is the dimension of the low dimensional space, $s$ is the number of negative samples, and $N$ is the number of points in the dataset.

One of the primary challenges faced with dimensionality reductions is the construction of the $k$-nearest neighbor graph, which from now on will be referred to as the $k$-NN graph. In t-SNE, vantage point trees \citep{10.5555/313559.313789} are used to construct the $k$-NN graph. Though this is an effective method for relatively large data sets, the performance greatly reduces as the dimensionality of the data increases.

The computational complexity of an exact $k$-NN can be modeled by the formula $O(N^2d)$, with $N$ being the total number of data points and $d$ the number of dimensions. This is considered very complex, so it is reduced using techniques for in-exact approximation such as space-partitioning trees \citep{10.1145/509907.509965,10.1145/997817.997857,10.5555/645925.671516}, and nearest neighbor exploration \citep{10.1145/1963405.1963487}. By partitioning the space, the data can be more easily organized into tree-structures while preserving the general structure of the data, with random projection trees \citep{10.1145/997817.997857} being used due to their high efficiency in nearest neighbor exploration. Locality sensitive hashing is then used to organize the data into “buckets” which contain data considered to be similar. Finally, the $k$-NN graph is refined and improved over a certain amount of iterations.

Going into greater detail, LargeVis begins by calculating the weights of the edges in the $k$-NN graph using the same method as t-SNE, modeled by Eqn. \ref{eqn:label1}. The graph is then made symmetric by finding the weight between all inputs, modeled by 

\begin{equation}
\label{eqn:label5}
 w_{ij}=\frac{p_{i|j}+p_{j|i}}{2N}
\end{equation}

It is at this point that LargeVis begins to use a unique probabilistic model for visualizing the $k$-NN graph. With a goal of maintaining similarity between vertices in low dimensional space, the probability of observing a binary edge $e_{ij}=1$ from a pair of vertices ($v_{i},v_{j}$) is modeled as follows,

\begin{equation}
\label{eqn:label6}
P(e_{ij}=1)=f(||\vec{y_{i}}-\vec{y_{j}}||)
\end{equation}
where $\vec{y_{i}}$ is the low dimensional embedding of $v_{i}$ and $\vec{y_{j}}$ of $v_{j}$. Here, $f()$ is any probabilistic function calculated with respect to the distance between the embedded vertices. The type of probabilistic function used can be varied to optimize results based on the structure of the data. Generalizing this equation for weighted edges, the probability of a weighted edge $e_{ij}=w_{ij}$ can be modeled by
\begin{equation}
\label{eqn:label7}
P(e_{ij}=w_{ij})=P(e_{ij}=1)^{w_{ij}}
\end{equation}

Given these, the probability of the weighted graph, $G=(V,E)$, is modelled as 

\begin{equation}
\label{eqn:label8}
O=\prod\limits_{(i,j)\epsilon E} p(e_{ij}=1)^{w_{ij}} \prod\limits_{(i,j)\epsilon \bar{E}} (1-p(e_{ij}=1))^{\gamma} 
\\ \propto \sum\limits_{(i,j)\epsilon E} w_{ij} \log{p(e_{ij}=1)}+\sum\limits_{(i,j)\epsilon \bar{E}} \gamma \log{(1-p(e_{ij}=1))}
\end{equation}
where $\bar{E}$ is the set of unobserved vertex pairs and $\gamma$ the collective weight of the negative edges. The first component of this formula shows the probability of observed edges, which when maximized, clusters similar points together in the low dimensional space. The second component models the probability of negative edges which, when maximized, ensures differing points stay far apart in the low dimensional space.

However, maximizing both of these equations is inefficient and computationally difficult. In order to mitigate this, negative edges are randomly sampled using a noisy distribution $P_{n}(j)$, with every vertex $i$ having randomly chosen vertices $j$, with $(i,j)$ representing the negative edges. With a noisy distribution of $P_{n}(j)\propto d_{j}^{0.75}$, where $d_{j}$ is the degree of the vertex $j$, and $M$ being the number of negative samples per positive edge, a more easily computable objective function can be found where

\begin{equation}
\label{eqn:label9}
O=\sum\limits_{(i,j)\epsilon E} w_{ij}(\log{p(e_{ij}=1)}+\sum\limits_{k=1} E_{j_{k}\sim P_{n}(j)}\gamma \log{(1-p(e_{ij_{k}}=1})))
\end{equation}

Finally, Eqn. \ref{eqn:label9} can be optimized using an asynchronous stochastic gradient descent, accelerating training and improving performance on sparse data sets. An innovative feature of LargeVis, in order to maintain the norms of the gradient, edges are randomly sampled \citep{10.1145/2736277.2741093} with a probability proportional to their weights and then used as binary edges in the stochastic gradient descent.

\subsubsection{UMAP \citep{mcinnes2018umap}}
\quad Uniform Manifold Approximation and Projection (UMAP) is a manifold learning algorithm used for the dimensionality reduction problem. The algorithm is rooted in Riemannian geometry and algebraic topology \citep{may1992simplicial}, allowing it to scale to much larger datasets, as compared to other algorithms, such as t-SNE.  UMAP is commonly used in a variety of fields, including bioinformatics \citep{bagger2016bloodspot,Becht298430,Diaz-Papkovich423632}, material science \citep{fuhrimann2018datadriven,Li2018}, and machine learning \citep{escolano2018selfattentive,blomqvist2018deep}. 

To give a general overview of the UMAP algorithm, local manifold approximations are combined to find their local fuzzy simplicial set representations \citep{goerss1999simplicial}. These representations are then used to construct a topological representation of the high dimensional data. Using this topological representation, a low dimensional representation of the data can be inferred. UMAP then optimizes the layout of the data representation in the low dimensional space in order to minimize the cross-entropy between the two topological representations.

Going into greater detail, we will discuss the specific computations in the UMAP algorithm. Being based heavily upon an in depth mathematical analysis of the problem, the motivation for certain steps in the algorithm can be found in Section 2 of \citep{mcinnes2018umap}.

To begin, UMAP constructs a weighted $k$-nearest neighbors graphs using an input high-dimensional data-set $X=\{x_{i},...,x_{n}\}$ and a dissimilarity measure $m : X \times X \to \mathbb{R}_{\geq 0} $. Using a given hyper-parameter $k$, the set of $k$-nearest neighbors $\{ x_{i1},...,x_{ik}\}$ is calculated according to the metric $m$ using nearest neighbor descent \citep{10.1145/1963405.1963487}.
\newline
Next, for every $x_{i}$, $\textit{p}_{i}$ and $\sigma_{i}$ are calculated such that
\begin{equation}
\label{eqn:label10}
\textit{p}_{i}=\textrm{min}\{ m(x_{i},x_{ij}) | 1 \leq j \leq k, m(x_{i},x_{ij}) > 0 \}
\end{equation}
and
\begin{equation}
\label{eqn:label10}
\sum^{k}\limits_{j=1}\textrm{exp}(\frac{-\textrm{max}(0,m(x_{i},x_{ij})-p_{i})}{\sigma_{i}})=\log_2 (k)
\end{equation}
To construct the weighted-directed graph $\bar{G}=(V,E,w)$, we use $V=X$, $E=\{ (x_{i},x_{ij}) | 1\leq j \leq k,1 \leq i \leq N\}$, and 
\begin{equation}
\label{eqn:label10}
w((x_{i},x_{ij}))= \textrm{exp}(\frac{-\textrm{max}(0,m(x_{i},x_{ij})-p_{i})}{\sigma_{i}}).
\end{equation}
Finally, the symmetric adjacency matrix $B$ of the undirected, weighted graph output for UMAP, $G$, can be defined by
\begin{equation}
\label{eqn:label10}
B=A+A^{\top}-A \circ A^{\top},
\end{equation}
where $\circ$ is the pointwise product and $A$ is the weighted adjacency matrix of $\bar{G}$

To calculate the graph layout for $G$, UMAP utilizes a force directed graph layout algorithm in the desired low dimensional space which, using a set of attractive forces on edges and a set of repulsive forces on vertices, converges the graph and makes it aesthetically pleasing. With hyper-parameters of $a$ and $b$, the attractive force between two vertices $i$ and $j$ at coordinates $y_{i}$ and $y_{j}$ is calculated as 
\begin{equation}
\label{eqn:label10}
\frac{-2ab||y_{i}-y_{j}||_{2}^{2(b-1)}}{1+||y_{i}-y_{j}||_{2}^{2}}w((x_{i},x_{j}))(y_{i}-y_{j})
\end{equation}
In order to reduce the computational cost, edges are sampled from every vertex that has an attractive force applied to it and then one is repulsed. The repulsive force is calculated as
\begin{equation}
\label{eqn:label10}
\frac{b}{(\epsilon+||y_{i}-y_{j}||_{2}^{2})(1+||y_{i}-y_{j}||_{2}^{2})}(1-w((x_{i},x_{j})))(y_{i}-y_{j})),
\end{equation}
where $\epsilon$ is an arbitrarily small number to prevent division by $0$. By default, it is set to $0.001$. $G$ is initialized to a spectral layout which allows for quicker convergence and improved stability of the algorithm.

\subsubsection{TriMap \citep{amid2019trimap}}
\quad The TriMap method is one of the most recent tool for dimensionality reduction. With the majority of dimensionality reduction techniques aiming to preserve local neighborhood structure, TriMap focuses precisely upon providing an algorithm which preserves the global structure of the data. One of the major differences in the implementation of the TriMap method is the use of triplets to create an embedding, as compared to the more common pairwise method used by other algorithms. Here, they are represented as $(i,j,k)$, with the interpretation being that point $i$ is closer to point $j$ than point $k$.

To give a general overview of the TriMap algorithm, using a low-dimensional representation of the data created using the PCA algorithm \citep{pca}, triplets from the high dimensional representation are used to refine the quality of the low dimensional representation.

Going into greater depth, TriMap begins by sampling a subset of triplets $\tau=\{(i,j,k)\}$ and giving each a weight $\omega_{ijk}\geq 0$, where large values signify that $(i,k)$ are more distant than $(i,j)$. Using this, the cost function for a triplet $(i,j,k)$ is defined as
\begin{equation}
\label{eqn:label10}
l_{ijk}:=\omega_{ijk}\frac{s(y_{i},y_{k})}{s(y_{i},y_{j})+s(y_{i},y_{k})} \textrm{,  where} \hspace{4pt} s(y_{i},y_{j})=(1+|y_{i}-y_{j}||^{2})^{-1}
\end{equation}
The unnormalized weight of a triplet in the high-dimensional space can be defined as
\begin{equation}
\label{eqn:label11}
\tilde{\omega}_{ijk}=exp(D_{ik}^{2}-D_{ij}^{2}) \leq 0
\end{equation}
where $D_{ij}$ and $D_{ik}$ are the distances between points $x_{i}$, $x_{j}$ and $x_{i}$, $x_{k}$ in the high dimensional space, respectively. To find the euclidean distance, it is scaled by 
\begin{equation}
\label{eqn:label12}
D^{2}_{ij}=\frac{||x_{i}-x_{j}||^{2}}{\sigma_{ij}}
\end{equation}
where $\sigma_{ij}$ is the product of $\sigma_{i}$ and $\sigma_{j}$. $\sigma_{i}$ is the average Euclidean distance between $x_{i}$ and the set of its nearest-neighbors, up to 6-th neighbors. In this way, depending upon the density of the data, $\sigma_{ij}$ will dynamically change the scaling. 

Finally, in order to improve the local accuracy of the algorithm, the previously calculated weights $\tilde{\omega}_{ijk}$ undergo a $\gamma$-scaled  log-transformation, accentuating smaller weights, and pushing all other points farther away. Therefore, the final weight is defined as follows.
\begin{equation}
\label{eqn:label13}
\omega_{ijk}=\zeta_{\gamma}(\frac{\tilde{\omega}_{ijk}}{W}+\delta) \hspace{4pt} \textrm{, where  }  \zeta_{\gamma}(u):=\log{1+\gamma u} 
\end{equation}
where $W=\textrm{max}_{(i',j',k')\epsilon\tau}\tilde{\omega}_{i'j'k'}$, $\gamma$ is a small constant, and $\delta>0$ is the scaling factor. Both $\delta$ and $\gamma$ can be set by the user.

Using all of this information, the embedding can be constructed. Considering a subset of all possible triplets $(i,j,k)$, where $j$ is one of $i$'s nearest neighbors and $k$ is one of $i$'s most distant neighbors, we now define our hyper-parameters. Where $m=10$ is the number of nearest neighbors for each point and $m'=5$ is the number of triplets sampled for each nearest neighbor. This leaves us with 50 nearest neighbor triplets per point. In order to account for noise in the data and provide a more robust model, we allow for $r=5$ randomly sampled, and ordered, triplets to be inserted, making the final count of triplets per point $m \times m' +r=55$. Utilizing the ANNOY library for the construction of the approximate nearest neighbors tree, the initial embedding is set to the PCA embedding and scaled to improve convergence. By doing this, the algorithm provides superior global structure preservation in the embedding. The final cost function for the Trimap algorithm is represented as
\begin{equation}
\label{eqn:label14}
l_{\textrm{ TriMap}}=\sum\limits_{(i,j,k)\epsilon\tau}l_{ijk}
\end{equation}

\subsection{Graph Visualization}
\quad Graphs are considered as high-dimensional data \citep{erdos1965dimension}. Thus, visualization of graphs is similar to vector data provided that the graph has a given structural information i.e., connectivity of vertices by edges. There are various methods in the literature for graph visualization. We select a few of them for our study based on the popularity, runtime and quality of the visualizations. We discuss our selected methods as follows.

\subsubsection{OpenOrd~\citep{martin2011openord}}
\quad The OpenOrd method is a high-performance tool for graph visualization that can generate layout of graphs having millions of vertices. This method employs the spring-electrical model in its underlying optimization function. It uses a multi-level approach for graph layout generation. At each level, a force-directed approach is used to generate layout at each step. A graph is coarsened to reduce the size. In the coarsening step, vertices are merged using the average link clustering approach. When it reaches to the coarsest level, coarsened steps are reversed back to gain the final layout of the graph. Prolongation step and layout generation are performed at tandem so that it consumes less time. It optimizes the following objective function using simulated annealing heuristic approach.

\begin{equation}
    \label{eqn:openord}
    min_{x_1, \ldots, x_n} \sum_i(\sum_j w_{ij}E(x_i, x_j)^2 + D_{x_i})
\end{equation}
In Eqn. \ref{eqn:openord}, $E(x_i, x_j)$ represents the attractive force between vertices $i$ and $j$ using Euclidean distance, $D_{x_i}$ represents the approximated repulsive force of vertex $i$, and $w_{ij}$ represents the weight of the edge between vertices $i$ and $j$.

OpenOrd has been implemented using parallel computing techniques. It also supports several other options e.g., edge cutting is to visualize the densely connected graph. OpenOrd has a few drawbacks which are listed below:
\begin{itemize}
    \item One of the drawbacks of this approach is that it uses an adjacency matrix for graph representation. Thus, it easily reaches the available memory limit for large graphs.
    \item Parallel version has another drawback, for multiple cores, several runs results in different layouts which do not preserve consistency as required in scientific computing.
\end{itemize}

\subsubsection{ForceAtlas2~\citep{jacomy2014forceatlas2}}
The ForceAtlas2 \citep{jacomy2014forceatlas2} method is an engineering advancement that integrates several force-directed methods in Gephi software \citep{bastian2009gephi} for general purpose network visualization. Authors represent attractive force $F_a$ and repulsive force $F_r$ of two vertices $i$ and $j$ in terms of Euclidean distance as follows: $F_a \propto E^a(i, j)$, and $F_r \propto E^{-r}(i, j')$, where $E^n(i, j)$ is defined as follows.
\begin{equation}
\label{eqn:forceatlas2}
    E^n(i, j) = \{(i_x-j_x)^2+(i_y-j_y)^2\}^{1/n}
\end{equation}
where, $i_x$ and $i_y$ are $(x, y)$-coordinates of vertex $i$. We can compute repulsive force of vertex $i$ with respect to vertex $j'$ similar to Eqn. \ref{eqn:forceatlas2}. Now, changing the value of $(a,r)$, we can generate different force-directed models~\citep{hu2005efficient,kobourov2012spring}. A repulsive force is computed when two vertices are not connected by an edge. For scale-free networks or graphs having low average degree, this computation can be asymptotically $O(n)$. Thus, the repulsive force computation is the most time-consuming part of force-directed models. In ForceAtlas2, the authors apply quadtree-based repulsive force approximation technique that reduce the overall runtime of the model~\citep{barnes1986hierarchical}. Specifically, this technique reduces the runtime of repulsive computation from $O(n)$ to $O(\log n)$. Though ForceAtlas2 supports multi-threading, its Java-based implementation consumes high amount of memory and is not very effective to utilize the memory bandwidth optimally. Nevertheless, it generates readable layout for both connected and disconnected graphs. Later, this model has been implemented in GPU to generate layout of graphs having millions of vertices \citep{brinkmann2017exploiting}.

\subsubsection{tsNET~\citep{kruiger2017graph}}
\quad The tsNET \citep{kruiger2017graph} is a version of the t-SNE~\citep{maaten2008visualizing} which has been formulated for graph layout problem. The attractive force is computed using a similar approach to t-SNE. First, the graph theoretic shortest path distance is calculated for all vertices. Then, the joint probability between the adjacent vertices is calculated using the t-SNE approach. For repulsive force, the logarithmic difference between non-adjacent vertices is computed for all vertices and then added to the average of it optimization function. The optimization function of tsNET is given below:

\begin{equation}
    C = \lambda_aC_{KL} + \lambda_c\frac{1}{N}\sum_{i,j\in V} \parallel y_i - y_j \parallel + \lambda_r\frac{1}{N}\sum_{i,j\in V}\log(||y_i-y_j||+\epsilon)
\end{equation}

This method can generate good quality graph visualization. However, it has some drawbacks such as this method runs slower compared to other state-of-the-art graph visualization methods. Thus, it cannot generate layouts of large graphs. Since, choosing a good value of perplexity parameter is difficult in t-SNE~\citep{wattenberg2016how}, the problem remains the same for tsNET. Sometimes, this method does not converge which results in an unreadable layout.
\begin{figure}[!htb]
    \centering
    \includegraphics[width=0.70\linewidth]{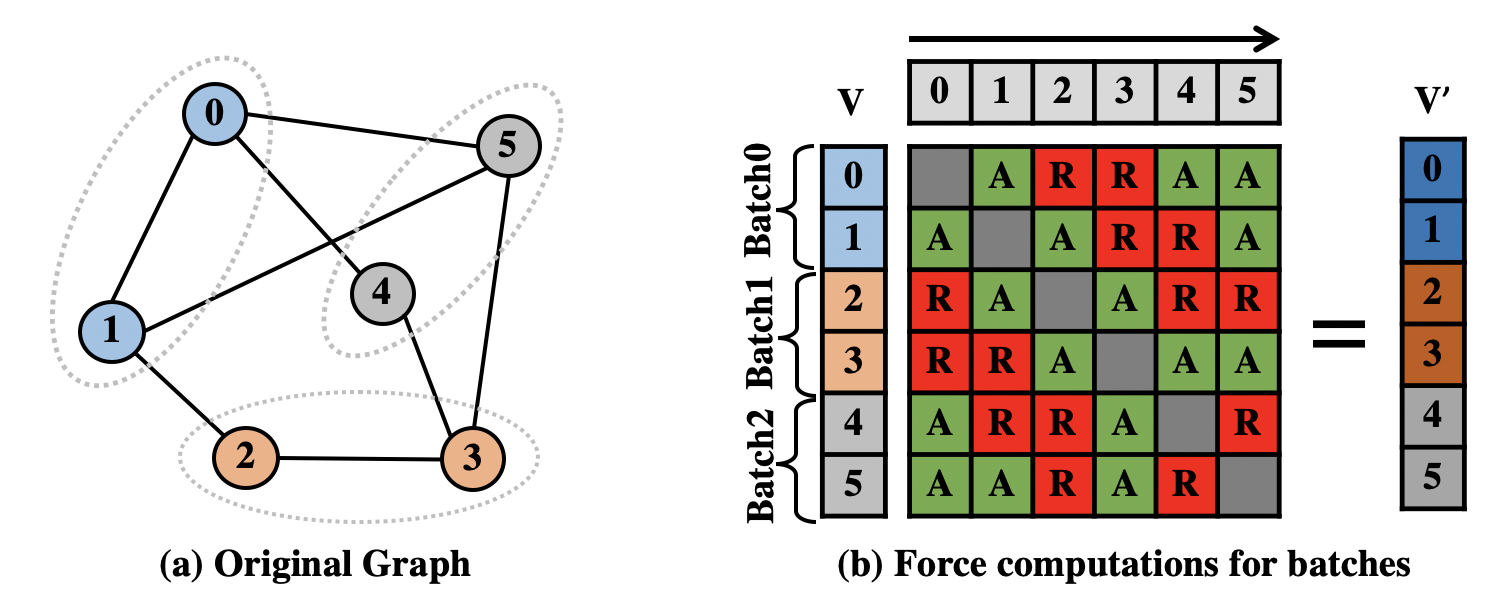}
    \caption{Force computations in BatchLayout: (a) The original graph where different batches are shown by a gray-dotted area. (b) Each batch contains two vertices and attractive and repulsive forces within a batch are computed in parallel and then coordinates are updated. $V$ and $V'$ represents vertex sets of before and after updating coordinates, respectively. `A' and `R' in each cell of the matrix represent attractive and repulsive force, respectively.}
    \label{fig:batchlayout}
\end{figure}
\subsubsection{BatchLayout~\citep{rahman2020batchlayout}}

BatchLayout is a very recent method that focuses on minimizing runtime and memory consumption while maintaining a good quality of the layout \citep{rahman2020batchlayout}. It employs the popular Fruchterman-Reingold \citep{fruchterman1991graph} spring-electrical models to preserve the quality of the layout, quad-tree data structure to reduce the runtime to $O(n\log n)$, and a mini-batch approach to obtain a faster parallel implementation. The original spring-electrical model has a sequential dependency similar to Stochastic Gradient Descent (SGD) which obstructs a massively parallel implementation. In the mini-batch approach, the attractive and repulsive forces are calculated for a subset of vertices in parallel. Then, the corresponding coordinates are updated based on the combined forces (see Fig. \ref{fig:batchlayout}). In Fig. \ref{fig:batchlayout}, there are three mini-batches such as $\{0, 1\}$, $\{2, 3\}$, and $\{4, 5\}$. We can compute the forces of verteces $0$ and $1$ in parallel as they are in the same batch. We do not consider updated coordinates of vertices $0$ and $1$ while computing forces with respect to these vertices. Then, we move to the next minibatch and compute forces of vertices $2$ and $3$. Now, we consider the updated coordinates of the previous batches i.e., $\{0, 1\}$. Since, the updated coordinates of these vertices are not considered while computing forces in a batch, the convergences of this approach takes higher iterations to converge. However, the parallel computing approach makes it significantly faster that easily compensate to generate good quality layout. Authors empirically show that the scalable parallel implementation of BatchLayout can catch the total number of converging iterations within a very short time compared to the sequential approach. In summary, the BatchLayout method can generate readable quality layout very fast compared to other methods.

\subsubsection{Other Graph Visualization Methods}
\quad Over the years, researchers have put efforts to reduce the runtime in force-directed models. Recently, a random sampling based repulsive force calculations approach has been introduced \citep{gove2019random}. Authors call it the Random Vertex Sampling (RVS) approach. The approach of this algorithm can be summarized as follows: select a subset of vertices $|V|^{\phi}$ to update the attractive force. Then, select $|V|^{\phi -1}$ randomly selected vertices for repulsive force calculations. Thus, the total time complexity will be $O(|V|^{\phi}.|V|^{\phi - 1}) = O(|V|)$, where, $\phi$ is a real number such that $0 < \phi < 1$. Authors proposed to update the force from left to right based on a sliding window approach so that each vertex gets a chance to be updated by repulsive forces. This method has another variation which runs $T-k$ iterations using random sampling approach and then runs last $k$ iterations using Barnes-Hut approach which produce good layout. Authors have tested their approach for graphs up to 100K vertices. Though this method introduces a fast sequential approach, it has some possible drawbacks which have not been considered in the model such as the randomly selected vertices can be neighbors (false negative) for repulsive force computation, and the suggested value of $\phi$ to be $\frac{3}{4}$ based on experiments, though graphs with a higher number of vertices will suffer as a result. It is obvious that for a small set of randomly selected vertices, the runtime for repulsive force calculations will always be low. However, it is also possible that the layout of such graph may not be readable. In the experimental results of the paper, the benchmark graph shows such negative results where RVS approach performs worse than Barnes-Hut approach.

There are various graph visualization softwares that can employ different graph visualization algorithms based on the users' expectation such as Gephi~\citep{bastian2009gephi}, GraphVis~\citep{ellson2001graphviz}, OGBD~\citep{chimani2013open}, etc. These tools provide either graphical user interface or command line interface while supporting  multi-core computations. They provide users a great advantage such as flexibility of choosing a graph visualization algorithm.

\section{Experiment}
\subsection{Experiment Setup}
\quad For our experiment, we analyze the runtime, memory performance, and aesthetic qualities of all four algorithms. In order to measure runtime, we use the \textit{time} function \footnote{\url{https://ss64.com/bash/time.html}} available in the bash shell, where the output ``real" time is measured. To measure the memory usage of each algorithm over time, we will utilize the memory\_profiler \footnote{\url{https://github.com/pythonprofilers/memory\_profiler}} package available for Python 3 . 
\newline
\null\quad For t-SNE, we use the original implementation distributed in the scikit-learn package \footnote{\url{https://scikit-learn.org/stable/modules/generated/sklearn.manifold.TSNE.html}} for Python 3. For LargeVis, we also use its original implementation in C++  \footnote{\url{https://github.com/lferry007/LargeVis}}. For UMAP, we use the distribution available on conda-forge, which is a redistribution of the original algorithm's GitHub  \footnote{\url{https://github.com/lmcinnes/umap}}. Finally, we use the TriMap implementation distributed on the Python Packed Index, which is sourced from the algorithms GitHub page \footnote{\url{https://github.com/eamid/trimap}}.
\newline
\null\quad We have run all tests on a system running RedHat Linux with an Intel(R) Xeon(R) CPU E5-2670 v3 CPU running at 2.30GHz. All of the tests have been run using the default parameters for each respective algorithm. For our vector data visualization experiments, we use the MNIST Digits dataset \footnote{\url{http://yann.lecun.com/exdb/mnist/}}, a standard dataset used in a large variety of machine learning applications, composed of labeled images of handwritten numbers, and the Fashion MNIST dataset \footnote{\url{https://github.com/zalandoresearch/fashion-mnist}}, a more modern and difficult dataset based on labeled images of clothing items.
\newline
\null\quad For the analysis of our results, we measure the runtime and memory usage of each algorithm. We will also consider the quality of the produced embedding, analyzing the quality of the clustering, which is the tendency for similar points to group together, the delineation between clusters, which is tendency for clusters to be separate from other clusters (i.e., not overlap), and the retention of the local and global structure of the embedded data. Our quantitative measures\footnote{\url{https://github.com/alexk101/dimensionality\_reduction\_measures}} for these qualities are their Davies-Bouldin and Silhouette scores, and  trustworthiness and global scores for their structures.

\null\quad The Davies-Bouldin score~\citep{4766909} measures the quality of the clustering through the average ratio of within-cluster distances to between-cluster distances, with the best score being 0. The similarity between clusters $R_{ij}$ is defined as 

\begin{equation}
    R_{ij}=\frac{s_{i}+s_{j}}{d_{ij}}
\end{equation}

where $s_{i}$ and $s_{j}$ are the average distance between each point of cluster $i$ and $j$ and the centroid, or diameter, of that cluster, and $d{ij}$ is the distance between cluster centroids $i$ and $j$. The final score is then calculated as

\begin{equation}
    DB = \frac{1}{k} \sum_{i=1}^k \max_{i \neq j} R_{ij}
\end{equation}

with $k$ being the total number of clusters.

\null\quad The silhouette score~\citep{ROUSSEEUW198753} measures the quality of the delineation between each point in each cluster on average, where positive values indicate superior delineation of values between clusters, values near 0 indicate overlapping clusters, and more negative values indicate points mapped to incorrect clusters. The score for one sample is calculated as

\begin{equation}
    s = \frac{b - a}{max(a, b)}
\end{equation}

where, $a$ is the average distance between the current point and all other points in the same class and $b$ is the average distance between the current point and all other points in the next nearest cluster. The overall score for an embedding containing multiple samples is calculated as 

\begin{equation}
    SC = \frac{\sum{i=1}^k s(i)}{n}
\end{equation}

where, $i$ is the current sample and $n$ is the total number of samples.

\null\quad The trustworthiness score is a measure that expresses how well the local structure of the data has been preserved by comparing the original dataset to the embedded dataset,  where the score ranges from 0 to 1, 1 being the best. The measure is calculated as

\begin{equation}
    T(k) = 1 - \frac{2}{nk (2n - 3k - 1)} \sum^n_{i=1}\sum_{j \in \mathcal{N}_{i}^{k}} \max(0, (r(i, j) - k))
\end{equation}

where for every sample $i$ up to the total number of samples $n$, $\mathcal{N}_{i}^{k}$ are its k-nearest neighbors in the embedding and each sample j is the $r(i,j)$-th nearest neighbor in the original space.

\null\quad The global score, defined within the implementation of TriMap as a quality measure, is the minimum reconstruction error of the original dataset through the use of a linear inverse map derived from PCA, where the score ranges from 0 to 1, 1 being the best. With $n$ data points such that ${x_{i}\epsilon\mathbb{R}^{m}}_{i=1}^{n}$, $X\epsilon\mathbb{R}^{m \times n}$ the ith column of the original dataset matrix corresponding to $x_{i}$, and $Y\epsilon\mathbb{R}^{d \times n}$ the embedded matrix corresponding to ${y_{i}\epsilon\mathbb{R}^{d}}_{i=1}^{n}$, the minimum reconstruction error (mre) for a given embedding is calculated as

\begin{equation}
    \varepsilon(Y|X):=\textrm{min}_{A\epsilon\mathbb{R}^{m \times n}} ||X- AY||_{F}^{2}
\end{equation}

where, $||\centerdot||_{F}$ is the Frobenius norm of the matrix. Using this mre, the global score is then formulated as

\begin{equation}
    GS(Y|X) := \textrm{exp}(-\frac{\varepsilon(Y|X)-\varepsilon_{pca}}{\varepsilon_{pca}})\epsilon[0,1]
\end{equation}

where $\varepsilon_{pca}:=\varepsilon(Y_{pca}|X)$, which is the ideal mre produced on the $X$ dataset using the PCA embedding method.

We use the implementation of these measures found in the scikit-learn package, except for the global score, which is a part of the TriMap method implementation and can be found on the paper's respective Github page.\footnote{\url{https://github.com/eamid/trimap}}$^,$\footnote{\url{https://scikit-learn.org/stable/modules/generated/sklearn.metrics.silhouette_score.html}}$^,$\footnote{\url{https://scikit-learn.org/stable/modules/generated/sklearn.metrics.davies_bouldin_score.html}}

\subsection{Results}

\subsection{Comparison of High-Dimensional Data Visualization Tools}

\begin{table}[!bth]
\centering
\caption{Actual (minutes:seconds) and theoretical runtimes for reducing the dimensionality of the MNIST datasets from 784 to 2 using four techniques (from left to right): t-SNE, UMAP, LargeVis, and Trimap}
\begin{tabular}{|l|c|c|c|c|}
\hline
\multicolumn{5}{|c|}{Visualization Program Runtimes}                                                                                     \\ \hline
                     & \multicolumn{1}{l|}{t-SNE} & \multicolumn{1}{l|}{UMAP} & \multicolumn{1}{l|}{LargeVis} & \multicolumn{1}{l|}{Trimap}       \\ \hline
MNIST Digits Actual  & 129:40.967                & 5:06.650                  & 9:35.643                      & \cellcolor[HTML]{C8C3BC}1:39.526  \\ \hline
Fashion MNIST Actual & 132:09.443                & 5:42.680                  & 9:57.080                      & \cellcolor[HTML]{C8C3BC}2:08.569  \\ \hline
Theoretical Runtime          & $O(N^{2})$                & $O(N^{1.14})$             & $O(smN)$                      & \cellcolor[HTML]{C8C3BC}$O(N)$    \\ \hline
\end{tabular}
\end{table}

\subsubsection{Runtime and Memory Consumption}
\quad To begin, we can see that all tested runtimes correspond with their respective theoretical runtimes, indicated in the bottom column of Table (1) for each algorithm. This tells us that the experiment ran as expected by the authors of all algorithms and is an accurate assessment of algorithmic runtime. Assessing the runtimes for all algorithms on the whole, we can see a general trend that the MNIST Digits dataset has a lower runtime for embedding than the Fashion MNIST dataset. This also follows our expectations, as the original intent of the Fashion MNIST dataset is intended and to be a more modern, drop-in replacement for the widely used, but dated, MNIST Digits dataset.
\newline
\null\quad With the original implementation of t-SNE taking the longest out of all the algorithms, running for more than 2 hours on both datasets, it does not provide a realistic runtime for datasets of great size. Though the version of t-SNE used in this experiment was the slowest of all, there are various other implementations of the algorithm that significantly improve its runtime, such as multi-core Barnes-Hut t-SNE, CUDA t-SNE, and anchored t-SNE, decreasing runtime by up to 700 times \citep{van2014accelerating,chan2018tsnecuda,fu2019atsne}.
\newline
\null\quad The next fastest algorithm for both data sets was LargeVis, with a runtime of approximately 10 minutes. Though this algorithm did not provide the quickest results, its runtime show that it can scale well to larger data-sets. Compared to t-SNE, the runtime for LargeVis is around 13 times lower. This is the only major jump that we see between the algorithms in terms of runtimes, with the remaining algorithms only providing minor speedups.
\newline
\null\quad UMAP was the second quickest algorithm tested, with its unique approach of applying topology mathematics to dimensional reduction resulting in  a runtime of less than 6 minutes for both data-sets. With this computational efficiency, UMAP can easily run on datasets over 70,000 data points in a realistic amount of time.
\newline
\null\quad Finally, TriMap was the fastest of all algorithms by a significant margin, with it outperforming all other algorithms in terms of runtime by at least 2 times and, when compared to t-SNE, up to 60 times. With runtimes of around 2 minutes on both data-sets and a linear computational complexity, it can certainly scale far beyond the tested 70,000 member data-sets it was tested on without any trouble.

\begin{table}[!tbh]
    \centering
    \caption{Memory consumption graphs for the two varieties of MNIST datasets, Fashion MNIST and the MNIST Digits. The values obtained show the memory consumption for reducing the dimensionality of the datasets from 784 to 2 using four techniques (from left to right): LargeVis,  TriMap, UMAP and t-SNE}
    \begin{tabular}{c c c c c c}
    \toprule
        & LargeVis & TriMap & UMAP & t-SNE \\
        \toprule
         \rotatebox{90}{MNIST Digits} &  \includegraphics[width=0.2\linewidth]{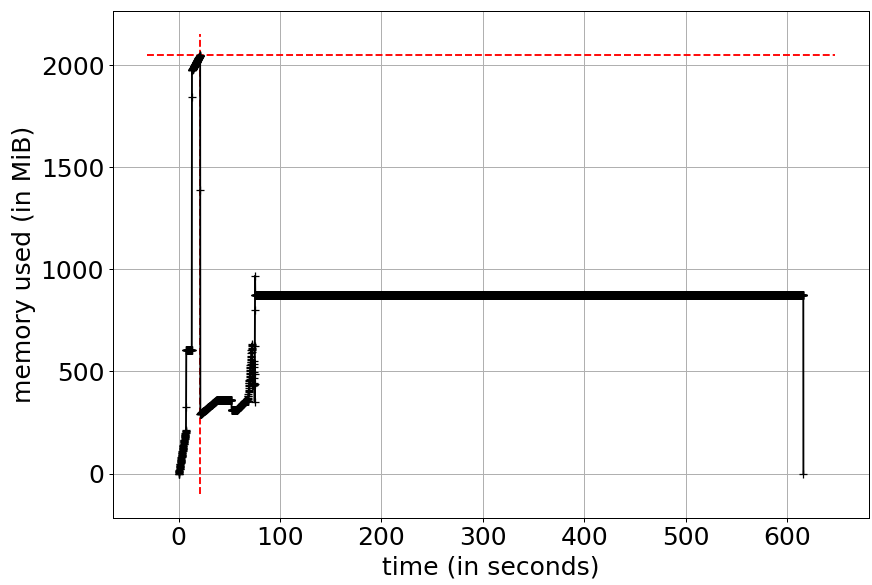} & \includegraphics[width=0.2\linewidth]{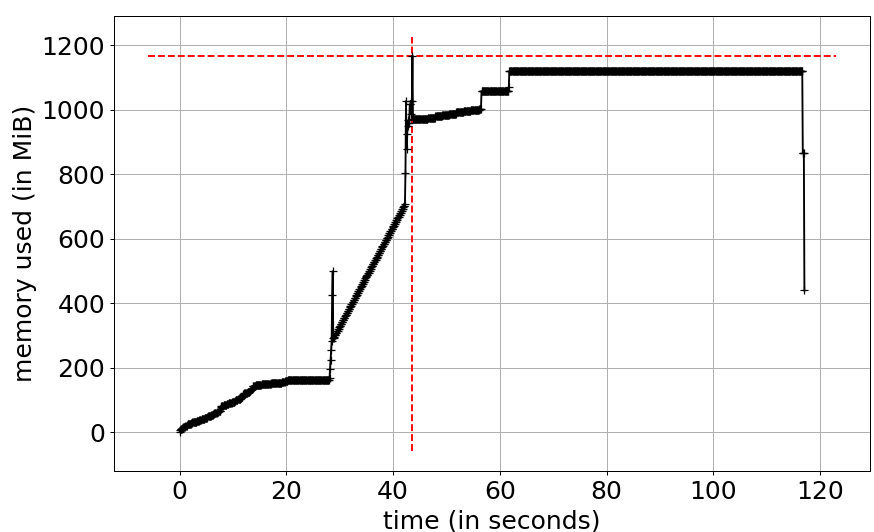} & \includegraphics[width=0.2\linewidth]{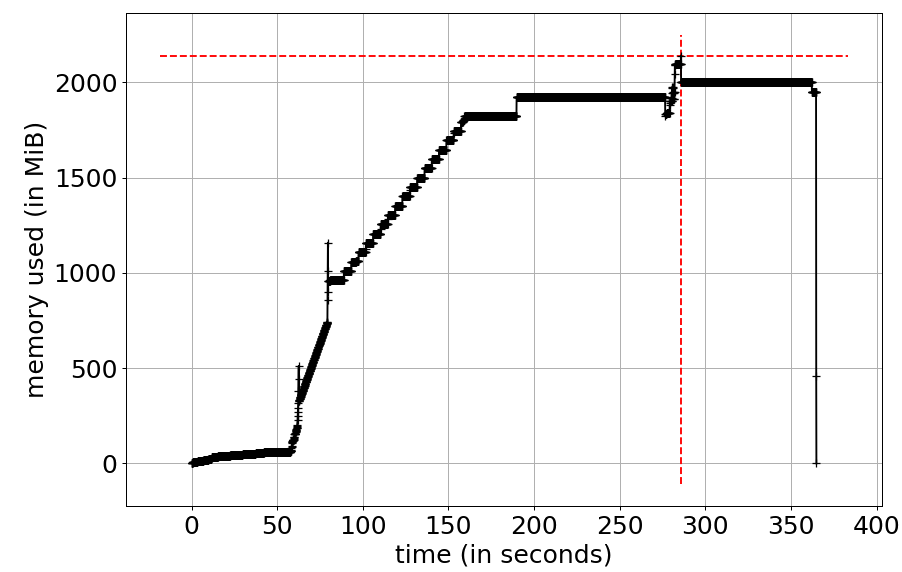} & \includegraphics[width=0.2\linewidth]{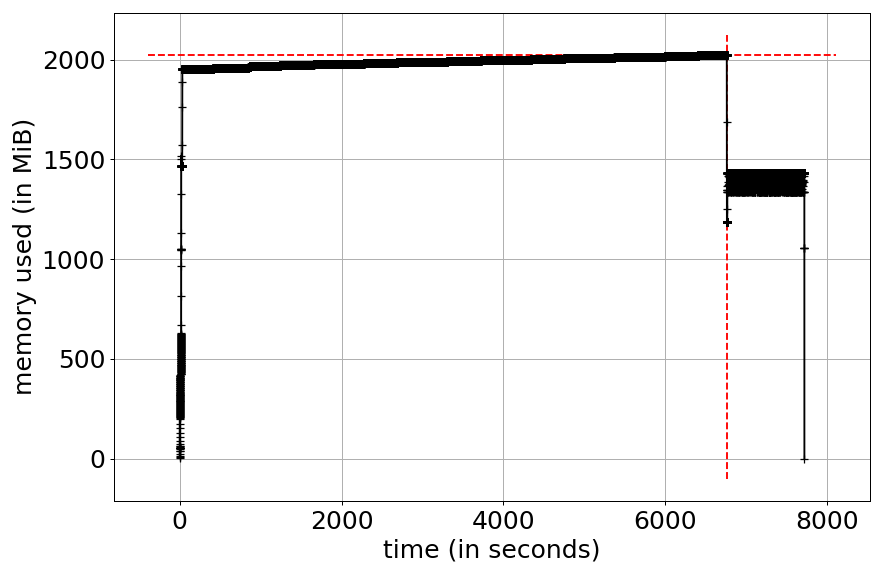} \\
         \midrule
         \rotatebox{90}{Fashion MNIST} &  \includegraphics[width=0.2\linewidth]{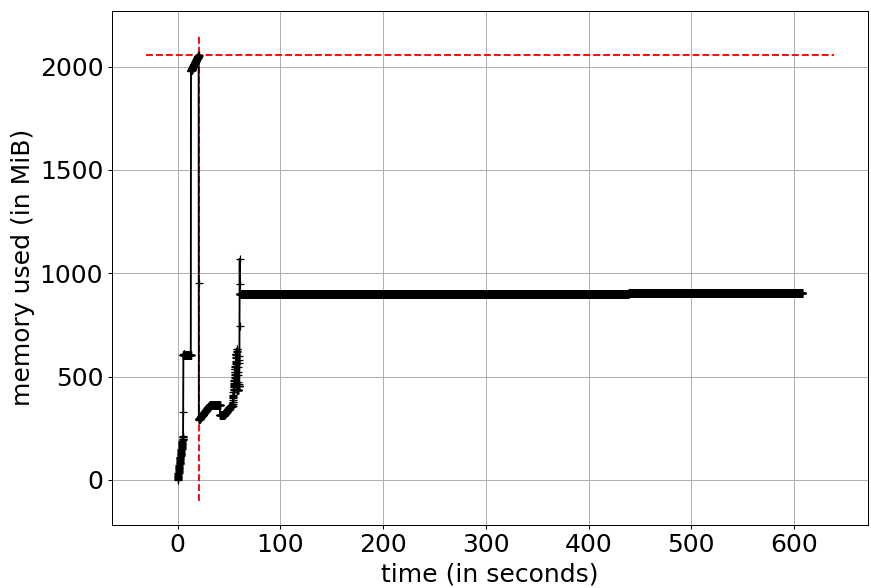} & \includegraphics[width=0.2\linewidth]{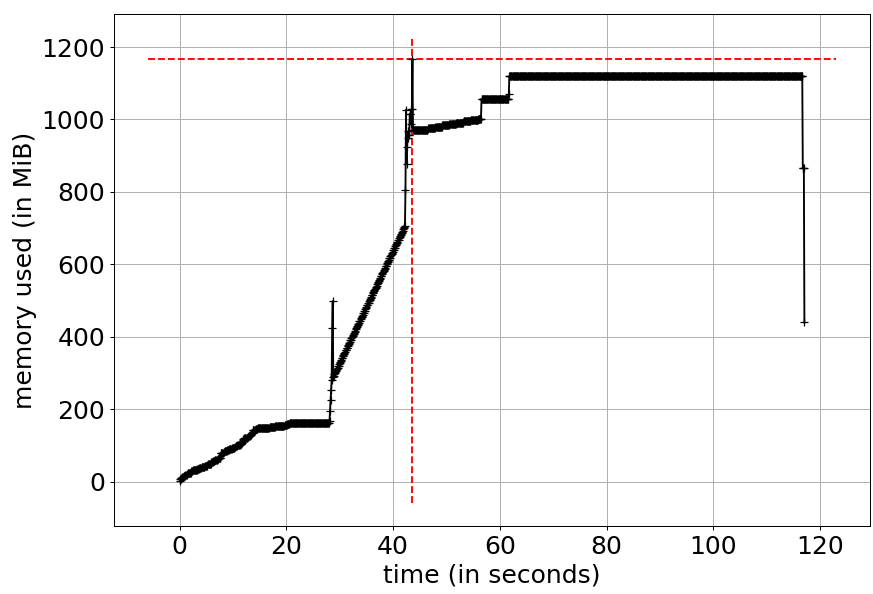} & \includegraphics[width=0.2\linewidth]{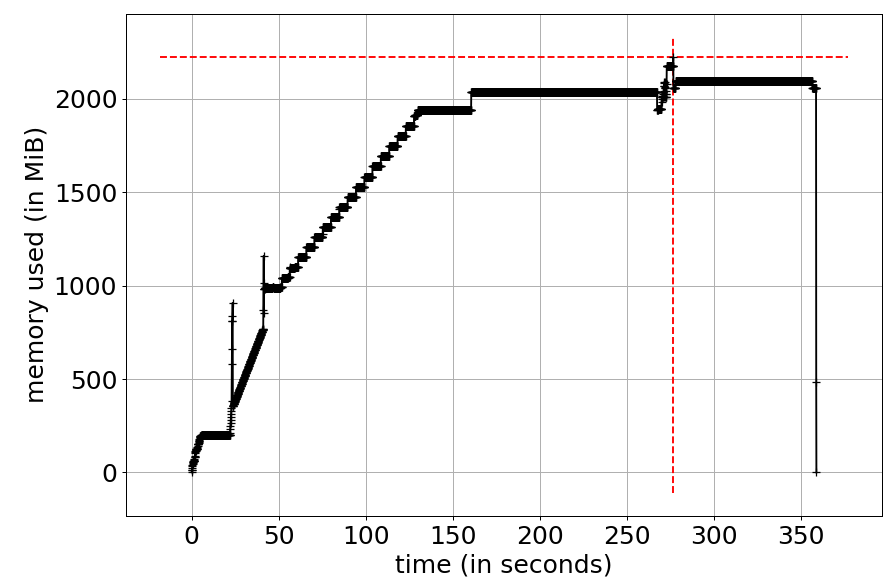} & \includegraphics[width=0.2\linewidth]{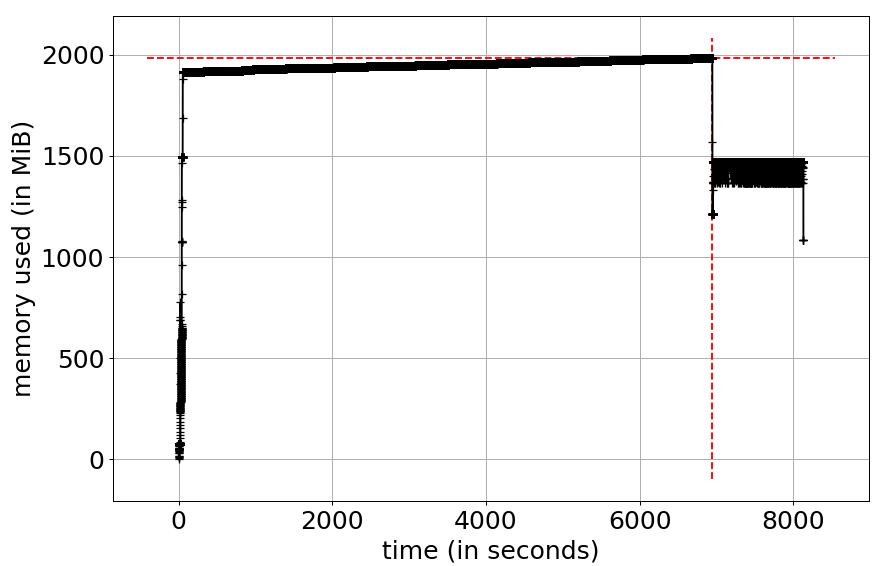} \\
         \bottomrule
    \end{tabular}
    \label{tab:mem}
\end{table}

\null\quad Looking at the memory used for t-SNE, we can see that for both data-sets, the memory usage holds at around 2 gigabytes for the majority of the runtime, only dropping down to approximately 1.5 gigabytes in the last fifth of the runtime. With this memory usage, t-SNE has the highest sustained memory consumption of all algorithms tested, making it less useful in systems with a more limited capacity, or for very large datasets. Though this is true, it also has the most predictable and stable memory usage, as compared to the other algorithms, plateauing almost immediately at a quite constant memory usage, until dropping down slightly towards the end of the test.
\newline
\null\quad For LargeVis, we see a peak memory usage of around 2 gigabytes occurring at the very start of the algorithm. Memory usage then quickly drops down and normalizes to a consistent value slightly less than 1 gigabyte. With a constant memory usage for the majority of its runtime, LargeVis is the second most stable algorithm according to memory usage.
\newline
\null\quad With a peak memory usage over 2 gigabytes, UMAP has the highest peak memory usage out of all algorithms tested. Furthermore, its memory usage is quite unstable, which several drops in usage paired with a series of sporadic climbs, only somewhat stabilizing at around half way through its runtime. 
\newline
\null\quad Finally, with a peak memory usage of about 1.2 gigabytes, TriMap is not only the quickest, but also the most memory efficient algorithm, allowing it to scale up to larger datasets without the need for a significantly larger amount of memory. However, the stability of its memory usage is among the worst of the algorithms tested, with it being comparable to that of UMAP with its frequent spikes in usage.

\begin{table}[!thb]
    \centering
    \caption{2D representations of the two varieties of MNIST datasets, Fashion MNIST and the MNIST Digits. The 2D representations are obtained by reducing the dimensionality of the datasets from 784 to 2 using four techniques (from left to right): LargeVis,  TriMap, UMAP and t-SNE}
    \begin{tabular}{c c c c c c}
    \toprule
        & LargeVis & TriMap & UMAP & t-SNE \\
        \toprule
         \rotatebox{90}{MNIST Digits} &  \includegraphics[width=0.2\linewidth]{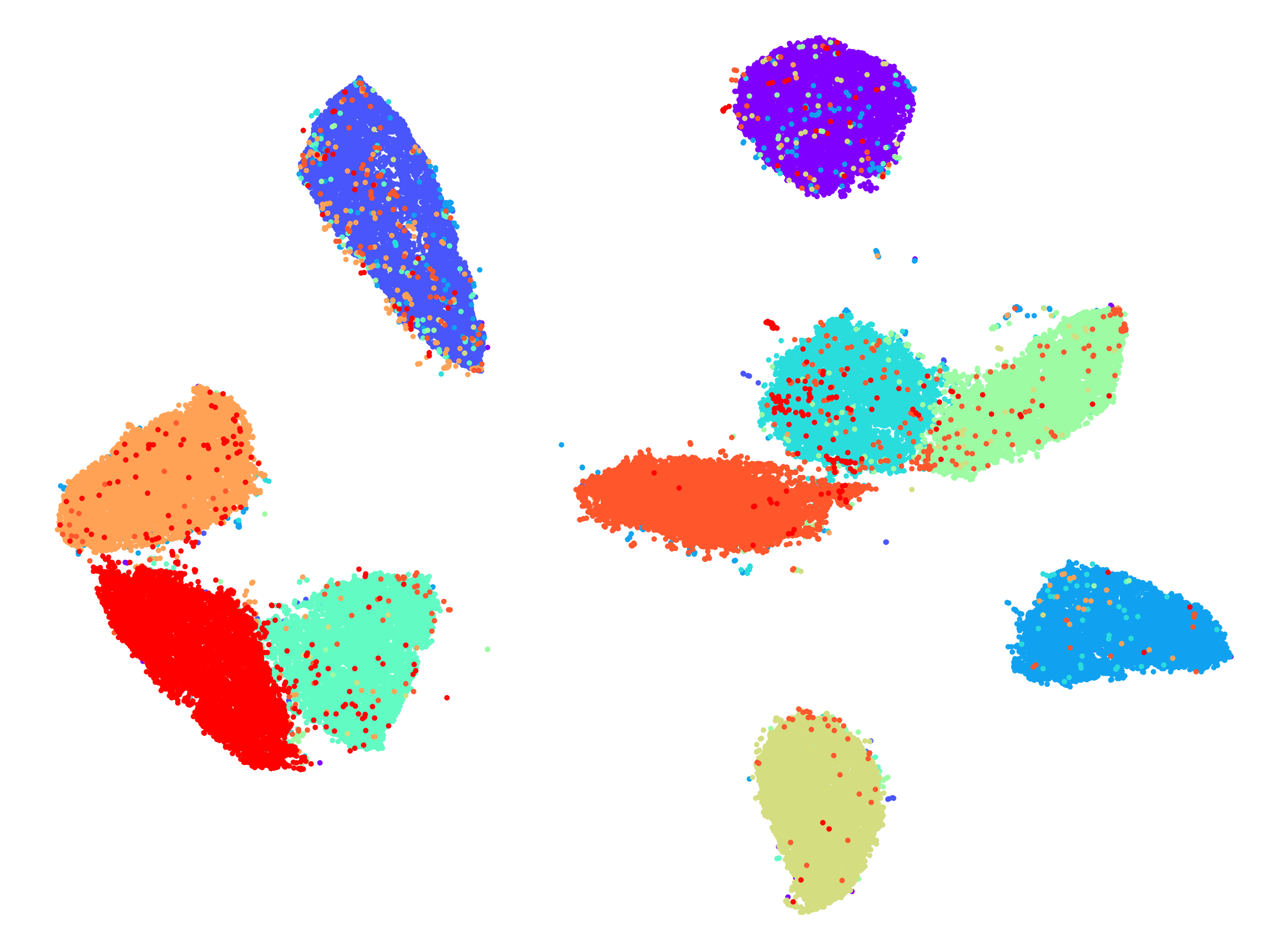} & \includegraphics[width=0.2\linewidth]{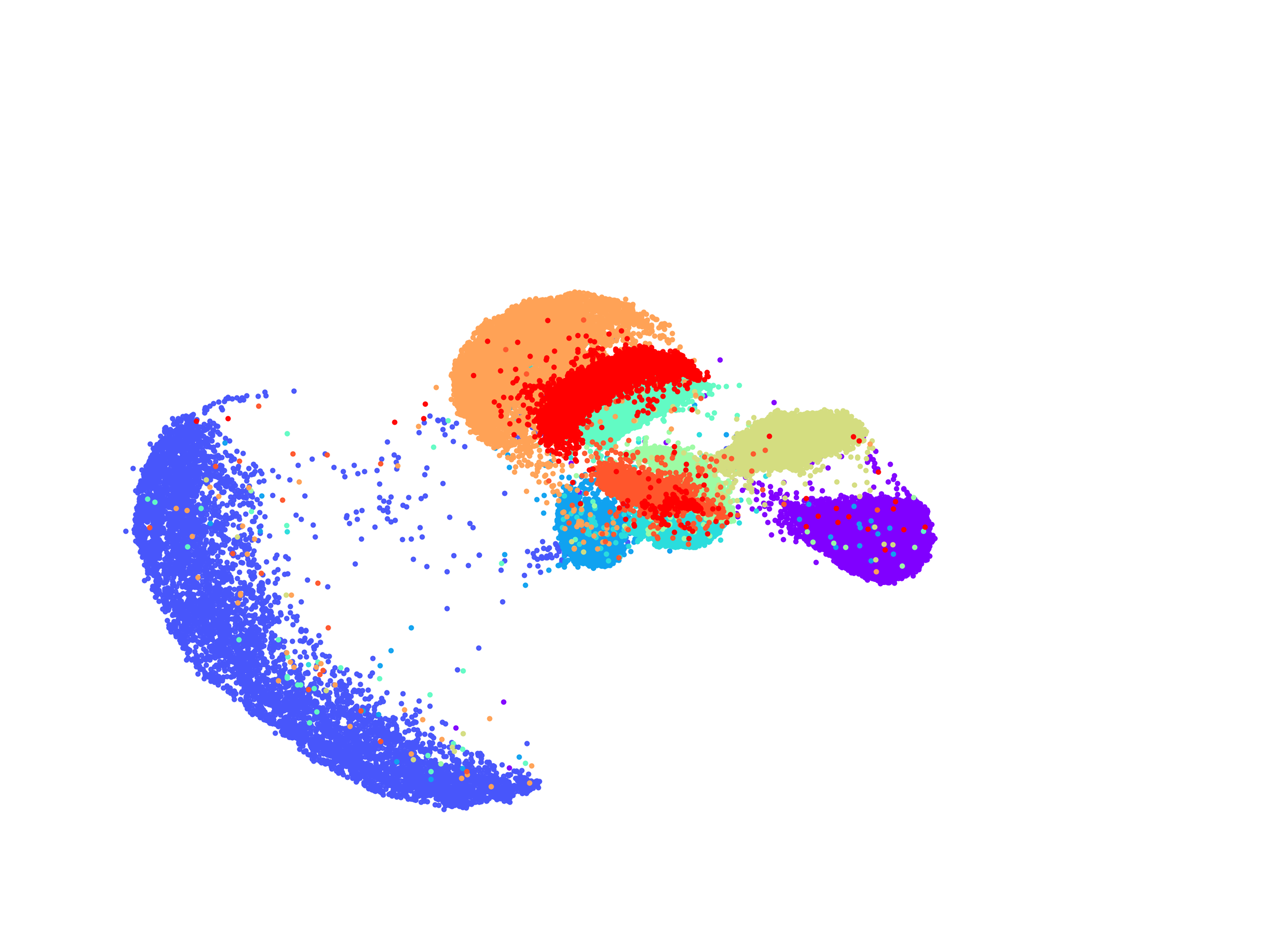} & \includegraphics[width=0.2\linewidth]{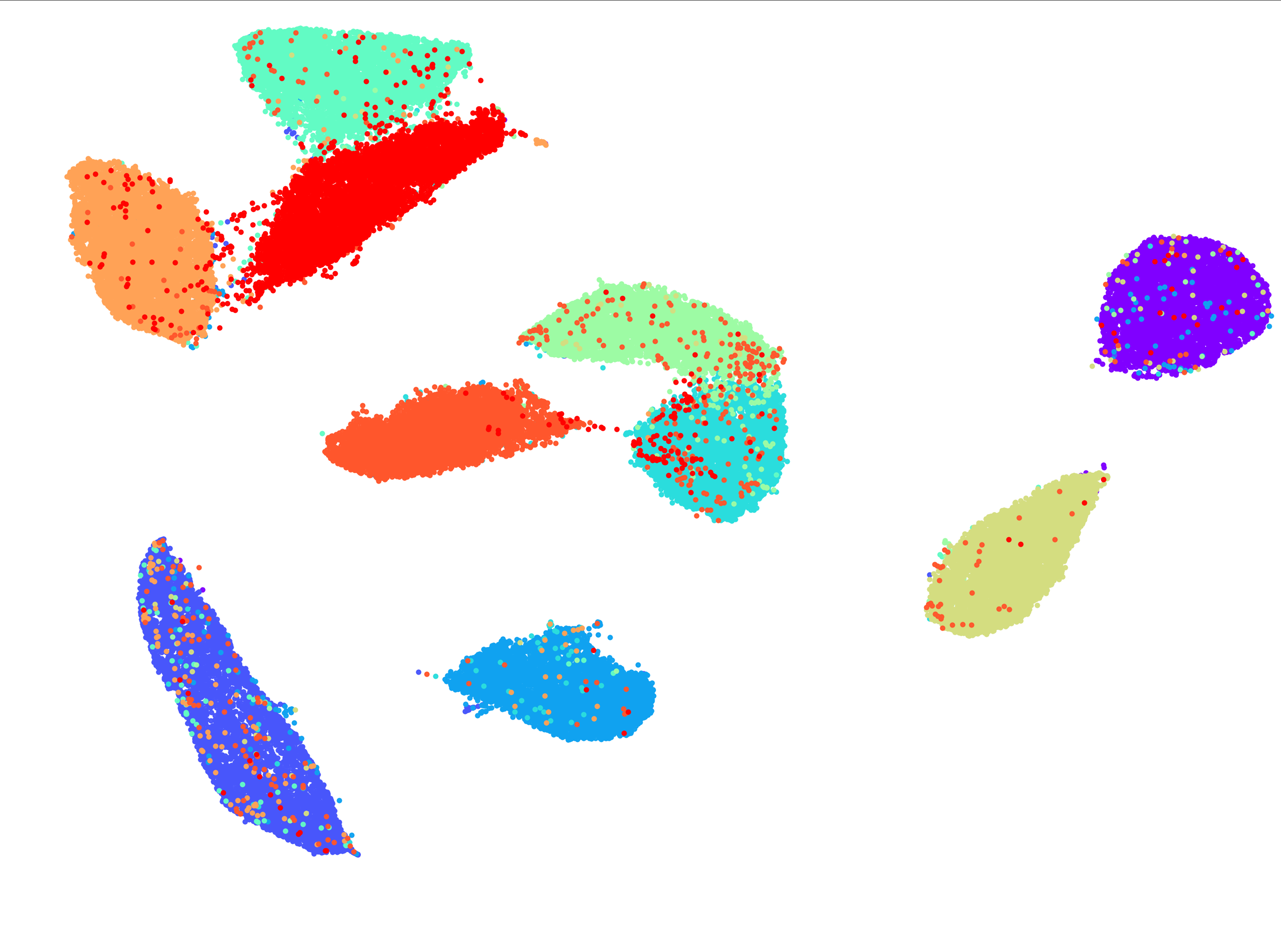} & \includegraphics[width=0.2\linewidth]{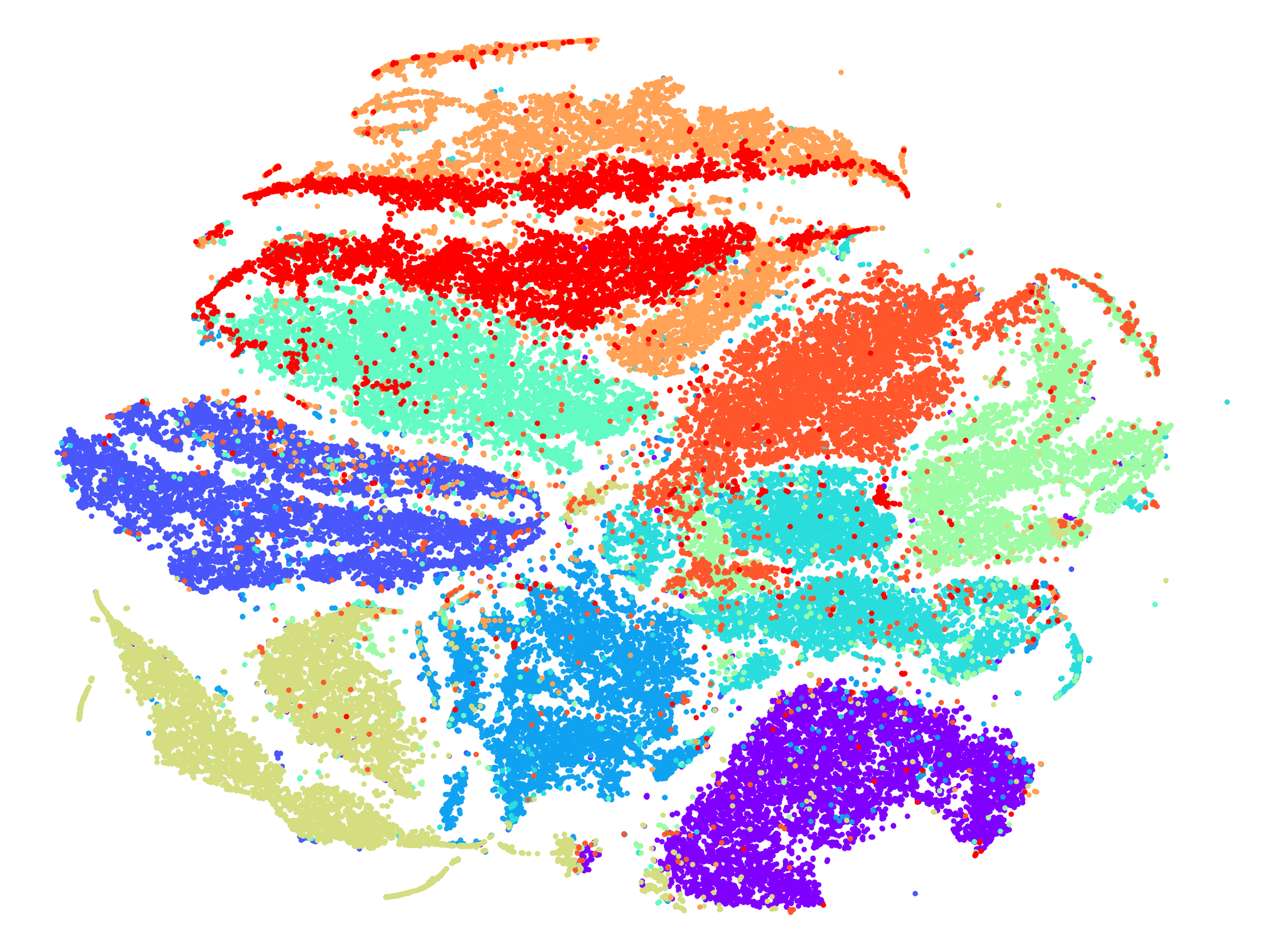} \\
         \midrule
         \rotatebox{90}{Fashion MNIST} &  \includegraphics[width=0.2\linewidth]{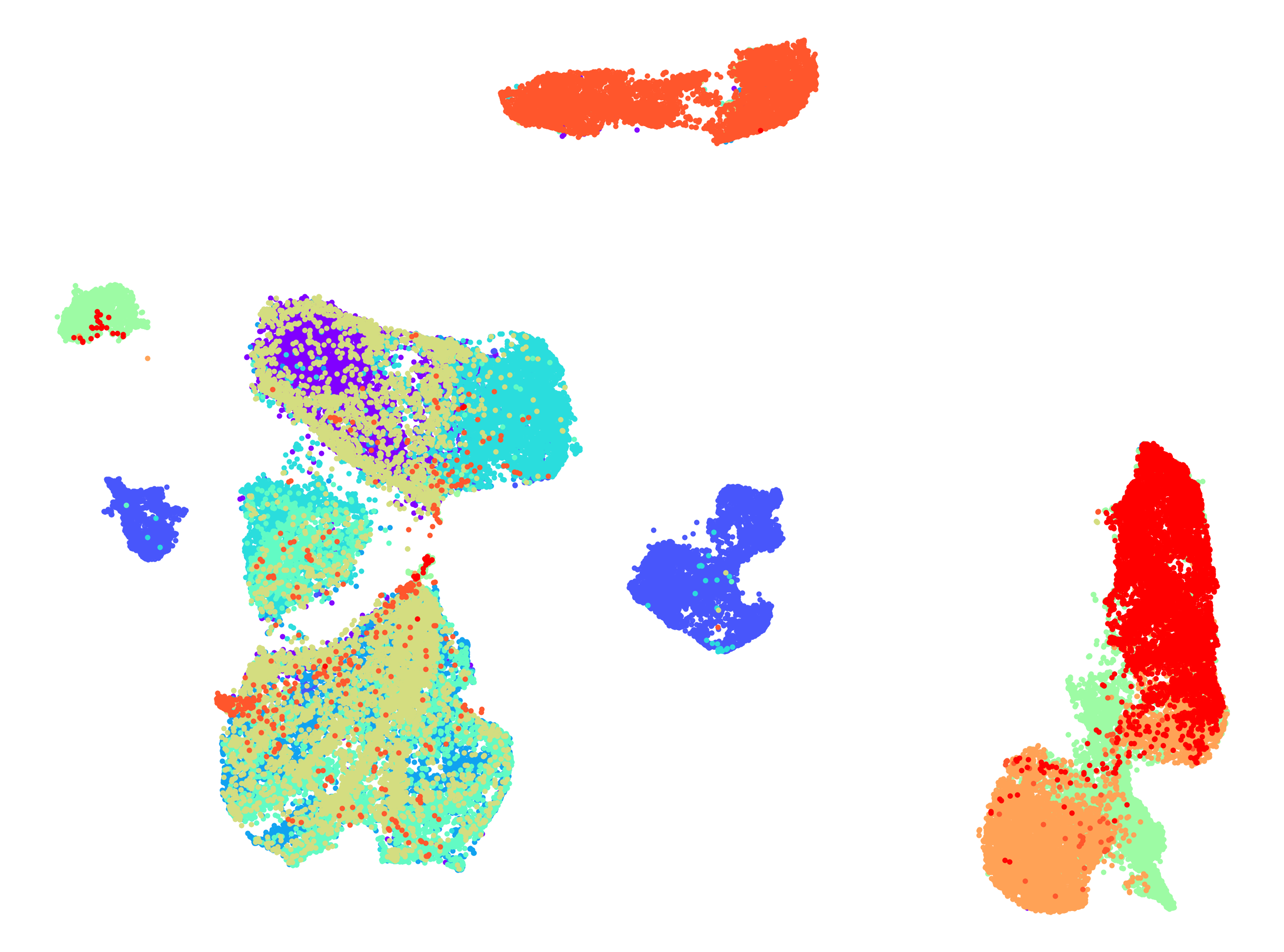} & \includegraphics[width=0.2\linewidth]{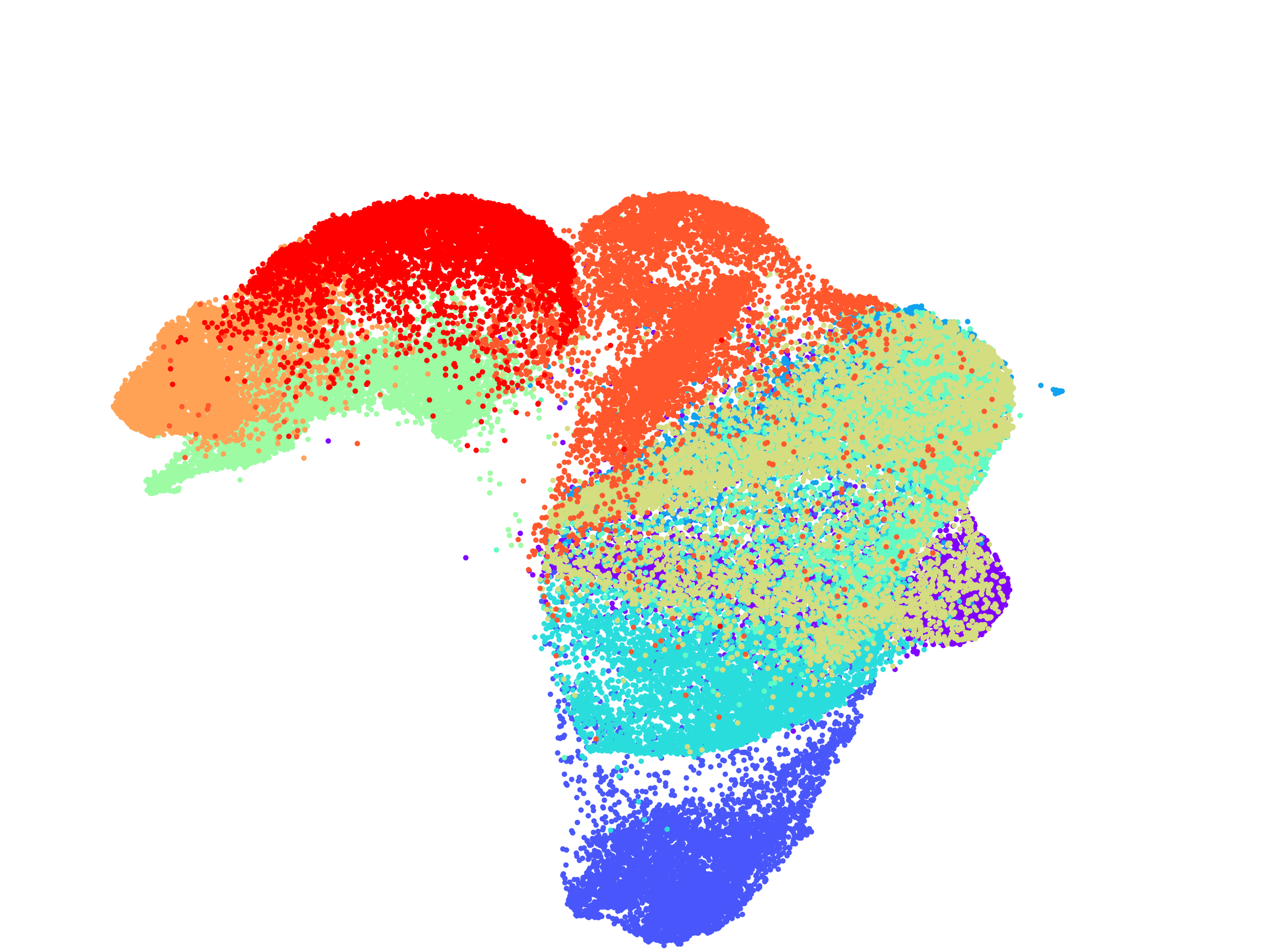} & \includegraphics[width=0.2\linewidth]{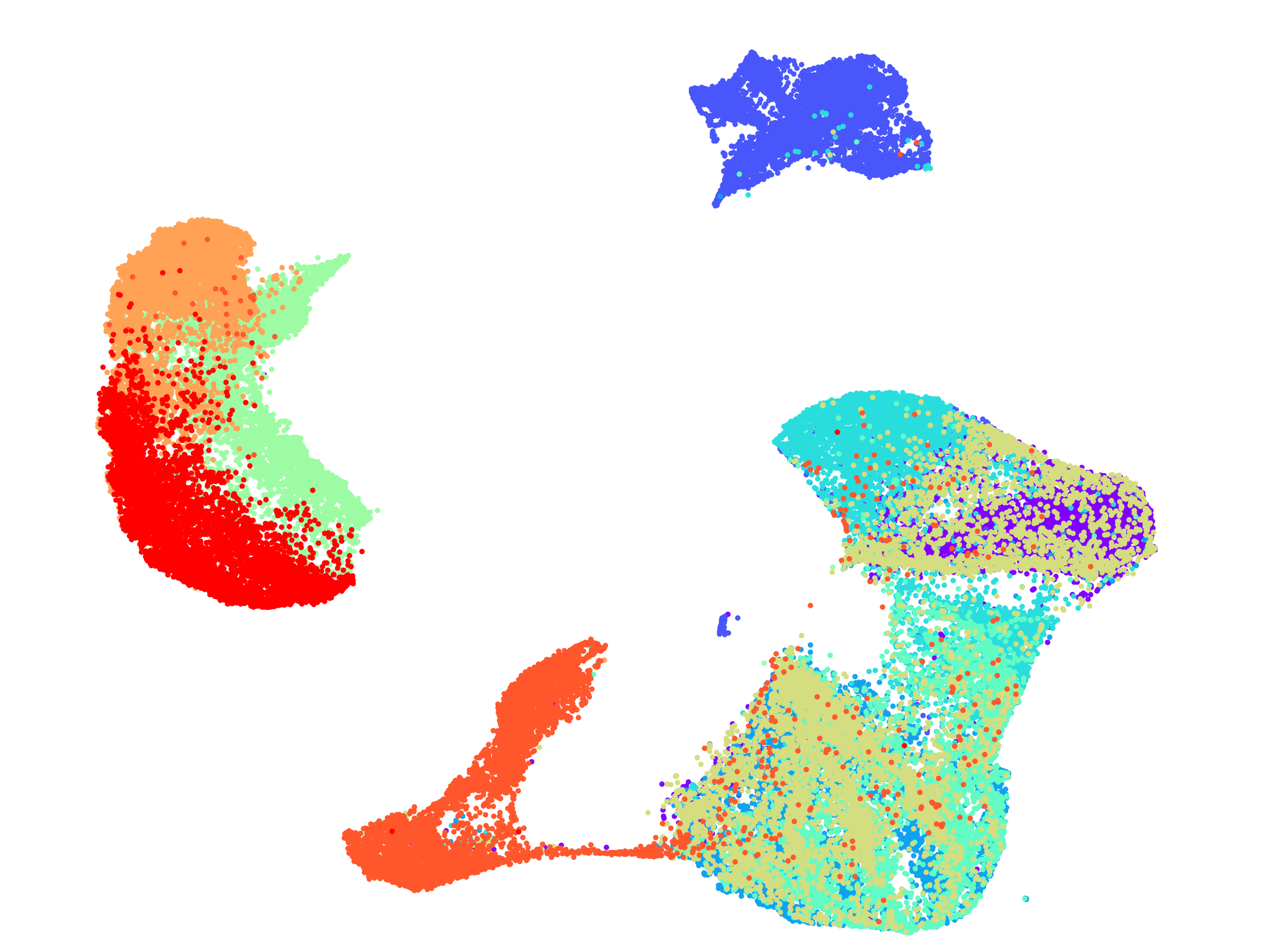} & \includegraphics[width=0.2\linewidth]{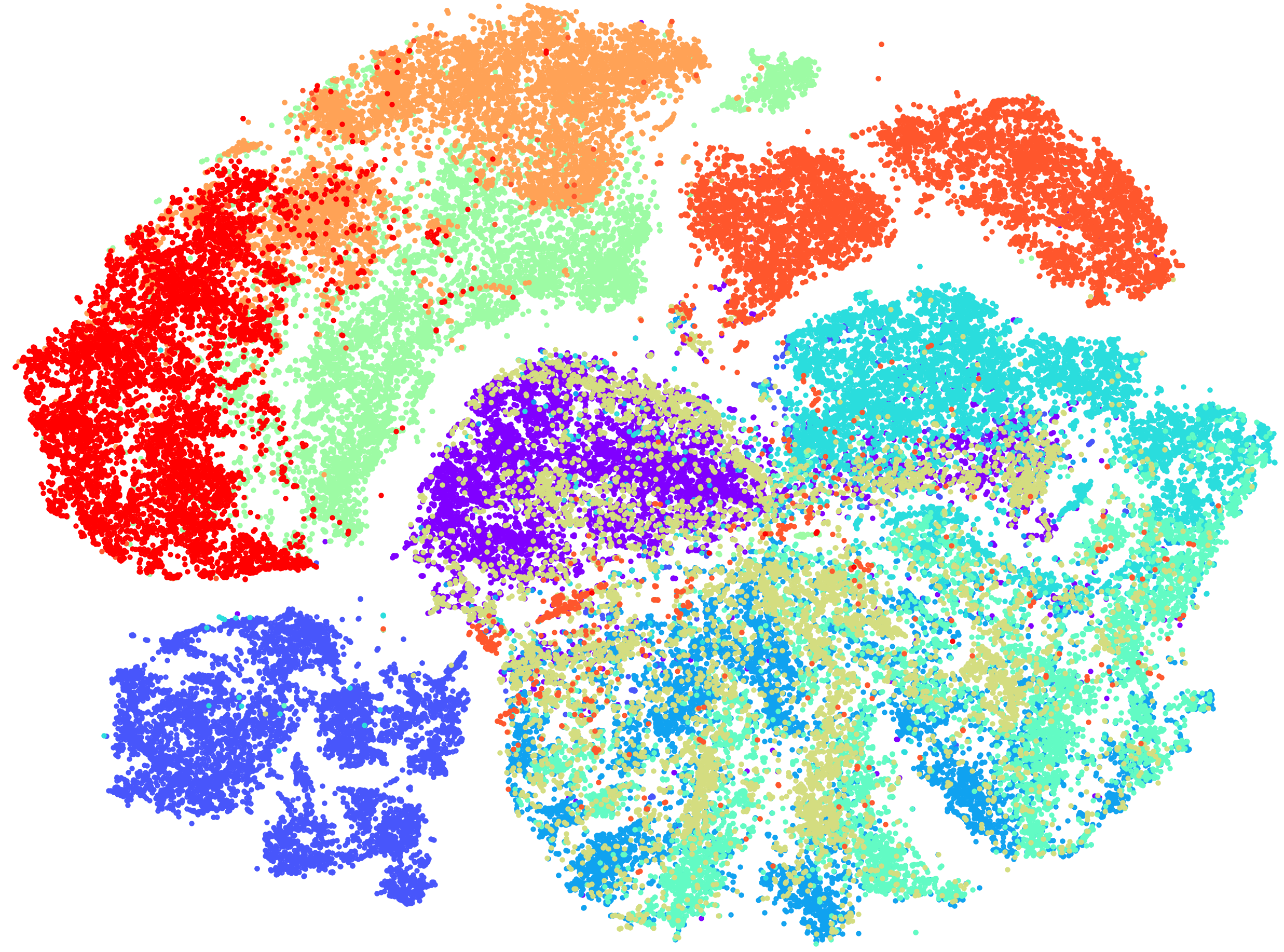} \\
         \bottomrule
    \end{tabular}
    \label{tab:embeddings}
\end{table}

\subsubsection{Qualitative Comparison}
\quad In Table \ref{tab:embeddings}, we present the final results of the each embedding algorithm. Comparing the quality of the embeddings between the datasets, we can see on the whole that the Fashion MNIST dataset is not as well separated between clusters as the MNIST Digits dataset, which is expected given the higher difficulty of the Fashion datasets. In terms of each specific algorithm, we can see similar qualities in the embeddings by LargeVis and UMAP, which tend to separate out the clusters more so than the other algorithms. This lends itself to the claim of global structure preservation presented by these algorithms. Looking at the results of TriMap, they are quite distinct from all the other algorithms, with clusters tending to concentrate about a centroid. Finally, we have t-SNE, which though the oldest of all the algorithms, seems to produce the most consistent results, with defined and well distributed clusters.

\begin{table}[!tbh]
\centering
\caption{Results of quantitative analysis of reduced dimensionality embeddings, focusing on the quality of certain clusters characteristics, as well as the preservation of neighborhood structure}
\begin{tabular}{|l|l|l|l|l|l|} 
\hline
\multicolumn{6}{|c|}{Visualization Program Quality Measures}                                                                                                                                            \\ 
\hline
                               &                 & t-SNE                                       & UMAP                                        & LargeVis & Trimap                                        \\ 
\hline
\multirow{4}{*}{MNIST Digits}  & Silhouette      & -0.04781                                    & {\cellcolor[rgb]{0.753,0.753,0.753}}0.11363 & 0.05754  & -0.01781                                      \\ 
\hhline{|~-----|}
                               & Davies-Bouldin  & 2.92073                                     & {\cellcolor[rgb]{0.753,0.753,0.753}}2.14496 & 78.92549 & 2.34602                                       \\ 
\hhline{|~-----|}
                               & Trustworthiness & 0.66168                                     & {\cellcolor[rgb]{0.753,0.753,0.753}}0.71668 & 0.69962  & 0.67551                                       \\ 
\hhline{|~-----|}
                               & Global Score    & {\cellcolor[rgb]{0.753,0.753,0.753}}0.95202 & 0.94188                                     & 0.93215  & 0.94389                                       \\ 
\hline
\multirow{4}{*}{Fashion MNIST} & Silhouette      & -0.11374                                    & -0.04145                                    & -0.11132 & {\cellcolor[rgb]{0.753,0.753,0.753}}-0.00672  \\ 
\hhline{|~-----|}
                               & Davies-Bouldin  & 6.38772                                     & 6.22085                                     & 11.15644 & {\cellcolor[rgb]{0.753,0.753,0.753}}2.97314   \\ 
\hhline{|~-----|}
                               & Trustworthiness & 0.66047                                     & {\cellcolor[rgb]{0.753,0.753,0.753}}0.74379 & 0.66200  & 0.70229                                       \\ 
\hhline{|~-----|}
                               & Global Score    & 0.80805                                     & 0.81065                                     & 0.68861  & {\cellcolor[rgb]{0.753,0.753,0.753}}0.83558   \\
\hline
\end{tabular}
\label{tab:emb_measures}
\end{table}
\subsubsection{Quantitative Comparison}
\quad To conduct a quantitative comparison of the final visualizations shown in Table \ref{tab:embeddings}, we will use the Davies-Bouldin and Silhouette scores as an assessment of clustering quality, and the Trustworthiness and Global Score metrics, as implemented in by the Trimap Python package\citep{amid2019trimap}. Of important note is the fact that the Davies-Bouldin and Silhouette scores do not assess whether or not the retrieval of information from the clusters is accurate, but rather, the inherent qualities of the clusters as they are in each embedding. We present these quantitative results in Table \ref{tab:emb_measures}. 
\newline
\null \quad Beginning with the results of our clustering analysis, we were able to find quite consistent results between datasets. For the older MNIST Digits dataset, UMAP provided the best scores for both the Silhouette and Davies-Bouldin metrics, outperforming the other algorithms by a fair margin. For the newer Fashion MNIST Digits dataset, Trimap provided the best scores for both the Silhouette and Davies-Bouldin metrics, with other algorithms performing similarly by the Silhouette score, but much worse by the Davies-Bouldin index.
\newline
\null \quad Highlighting the results of neighborhood structure preservation, we found that, for both datasets, UMAP was the most adept and maintaining the local structure of our datasets, as evidenced by its maximal Trustworthiness scores. However, in terms of the preservation of global structure, we saw superior performance from Trimap on the more challenging Fashion MNIST dataset, while for the older MNIST Digits dataset, global scores were all within 2 percentage points of one another, with t-SNE edging out Trimap just barely. 

\subsection{Comparison of Graph Visualization Tools}

\begin{table}[!ht]
\centering
\caption{Runtimes (seconds) of different graph visualization methods for different graphs shown in Table \ref{tab:layouts}. BL - BatchLayout, BLBH - Barnes-Hut based BatchLayoout, FA2BH - Barnes-Hut based ForceAtlas2, OO - OpenOrd.}
\arrayrulecolor{black}
\begin{tabular}{|c|c|c|c|c|c|c|} 
\arrayrulecolor{black}\hline
Graphs  & \#Vertices      & BL            & BLBH           & FA2BH & OO    & tsNET       \\ 
\hline
grid2\_dual & 3,136 & \textbf{3.31} & 4.21           & 6.30  & 5.90  & 632.92      \\\hline 

3elt\_dual  & 9,000 & 16.67         & \textbf{9.25}  & 19.40 & 16.50 & 3,396.66    \\ \hline

tube2      &    21,498 & 81.23         & \textbf{35.17} & 60.27 & 43.69 & 32,162.03   \\ \hline

finance256  &   37,376   & 234.66        & \textbf{47.57} & 81.54 & 69.19 & 163,532.71  \\ 
\arrayrulecolor{black}\hline
\multicolumn{2}{|c!{\color{black}\vrule}}{Theoretical Runtime} &   $O(n^2)$    &   $O(n\log n)$  &   $O(n\log n)$  &   $O(n^2)$     &   $O(n^2)$ \\ \hline
\end{tabular}
\label{tab:graphruntime}
\end{table}

\subsubsection{Runtime and Memory Consumption}
With the emergence of bigdata, large-scale graph drawing has become an important aspect. To big graphs having the number of vertices more than one million, there exist several methods in the literature. Gephi has ForceAtlas2 model which can utilize the available cores in machine. OpenOrd also runs in parallel. Recently, BatchLayout method has been proposed in the literature to effectively utilize cache memory and speedup force computation in shared memory architecture. On the other hand, tsNET generates good quality layout but runs sequentially. Thus, it can not generate layouts for large graphs. We report the runtime of four methods in Table \ref{tab:graphruntime} along with their theoretical complexity. We observe that one version of BatchLayout always runs faster than other methods. For a graph having 9,000 vertices, tsNET takes around an hour to generate a layout whereas BatchLayout Barnes-Hut (BLBH) version can generate good quality layout within 10 seconds. For finance256 graph, tsNET took almost 2 days to generate a layout whereas BLBH can generate a readable layout within a minute. BatchLayout and OpenOrd consume less memory than Gephi's ForceAtlas2 and tsNET. BatchLayout uses compressed sparse row (CSR) representation of graph structure which significantly reduces the memory consumption cost for scale-free networks. On the other hand, Gephi's
GML format consumes significant memory due to storing different node-link related information. Specifically, for grid2\_dual graph, BatchLayout, OpenOrd, ForceAtlas2, and tsNET consume 2.6MB, 17MB, 632MB, and 710MB, respectively. Thus, for low memory consumption and runtime, BatchLayout is a probable choice to generate graph layouts.

\begin{table}[!bht]
    \centering
    \label{tab:layouts}
    \caption{2D layouts of four representative graphs with each graph shown in a row of the table.
    2D layouts are obtained by embedding graphs directly on 2D space using five techniques (from left to right): BatchLayout,  BatchLayoutBH, ForceAtlas2BH, OpenOrd and tsNET. BatchLayoutBH - Barnes-Hut approach of BatchLayout, ForceAtlas2BH - Barnes-Hut approach of ForceAtlas2.}
    \begin{tabular}{c c c c c c}
    \toprule
        & BatchLayout & BatchLayoutBH & ForceAtlas2BH & OpenOrd &  tsNET \\
        \toprule
         \rotatebox{90}{grid2\_dual} &  \includegraphics[width=0.16\linewidth]{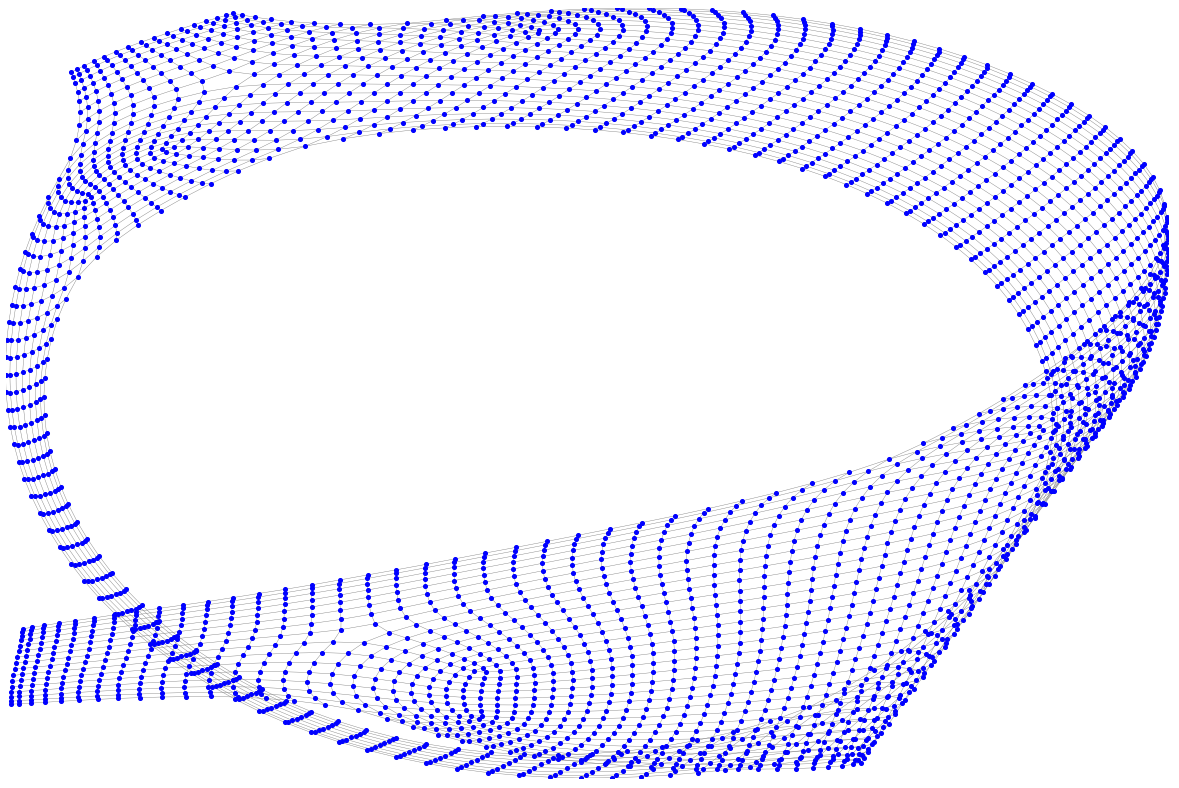} & \includegraphics[width=0.16\linewidth]{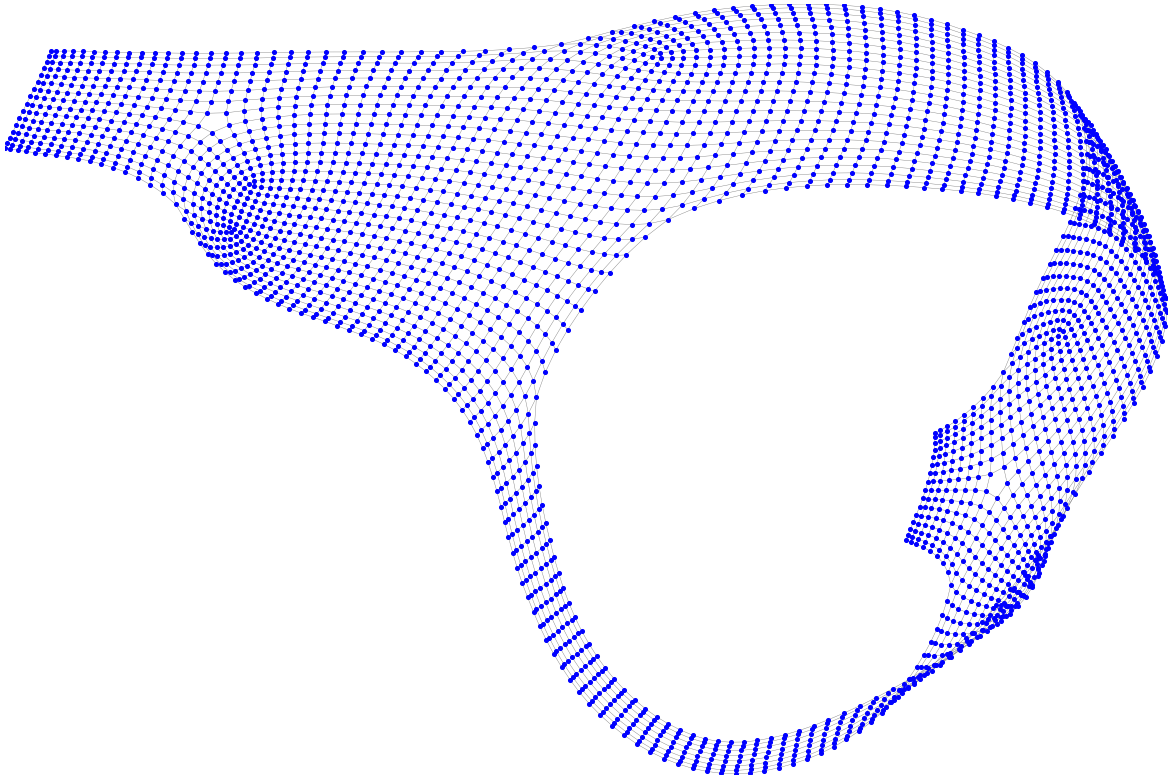} & \includegraphics[width=0.16\linewidth]{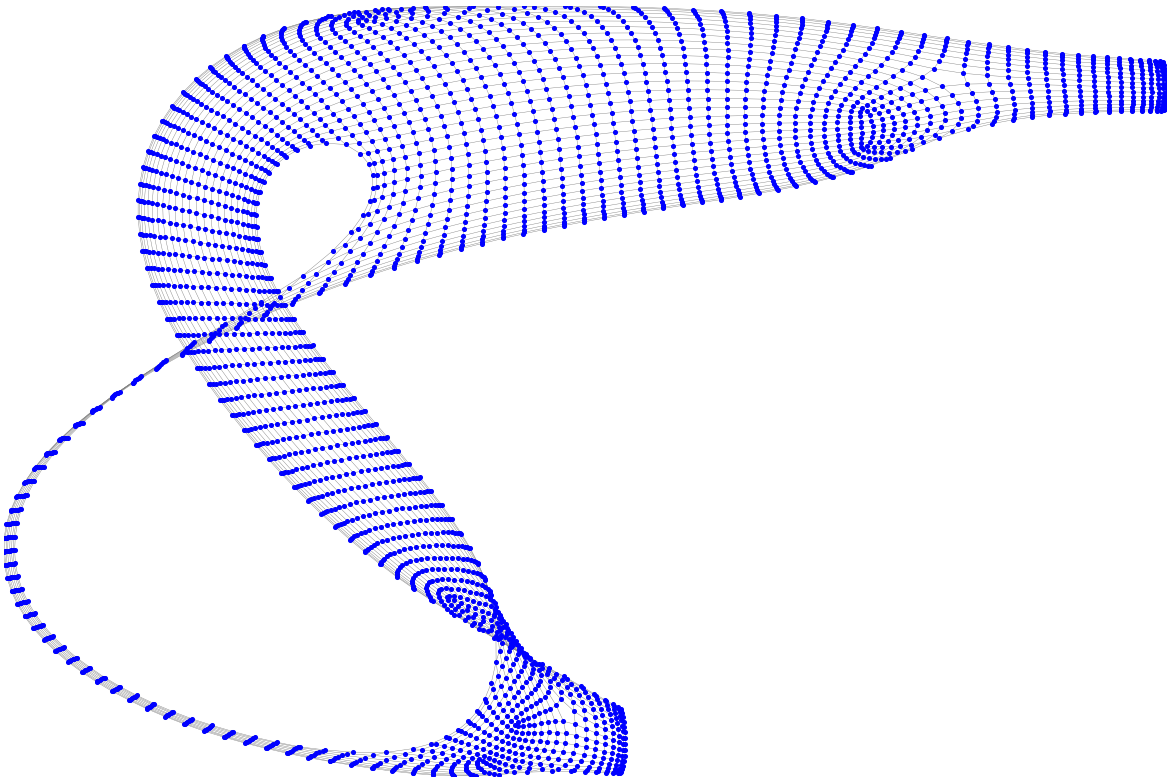} & \includegraphics[width=0.16\linewidth]{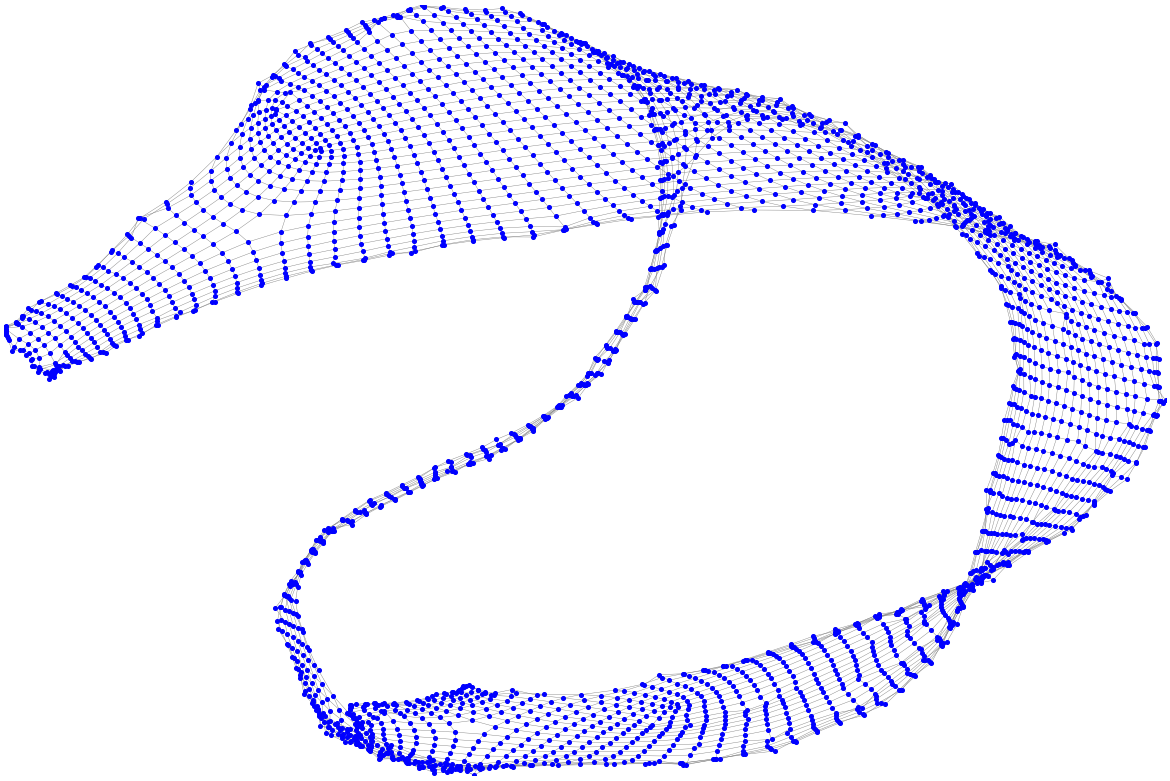} &
         \includegraphics[width=0.16\linewidth]{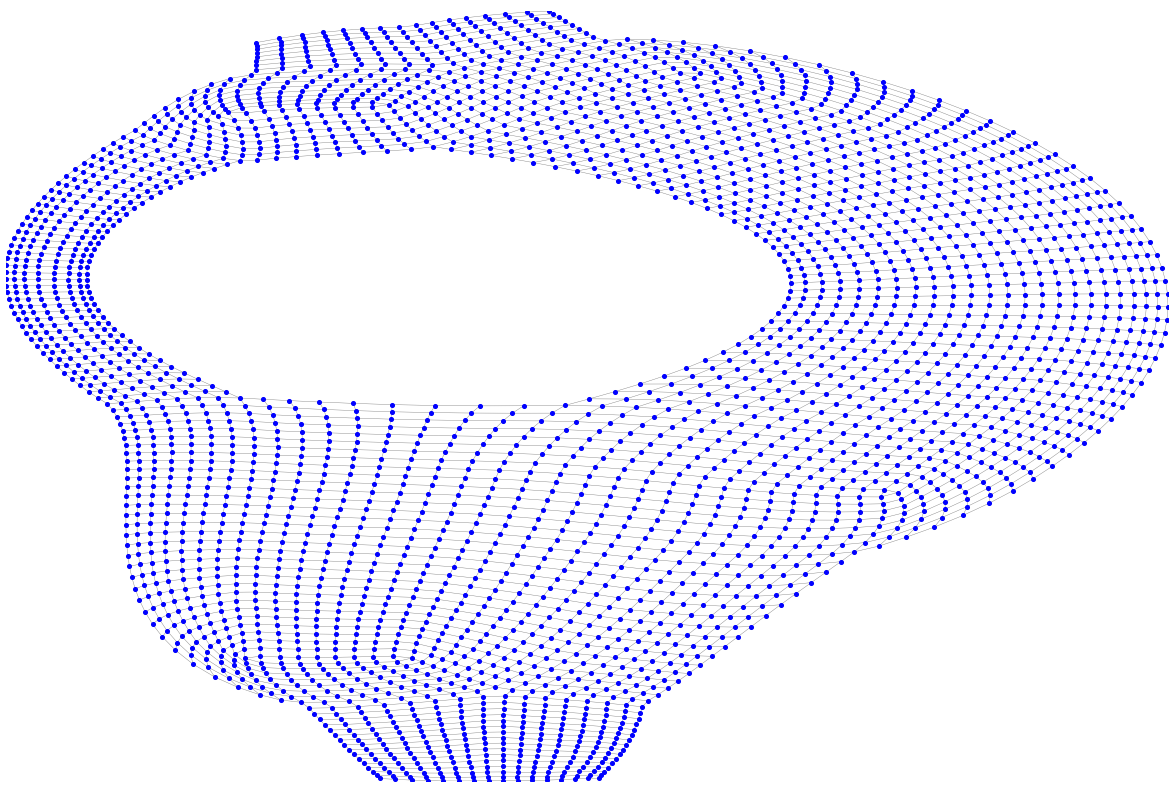} \\
         \midrule
         \rotatebox{90}{3elt\_dual} &  \includegraphics[width=0.16\linewidth]{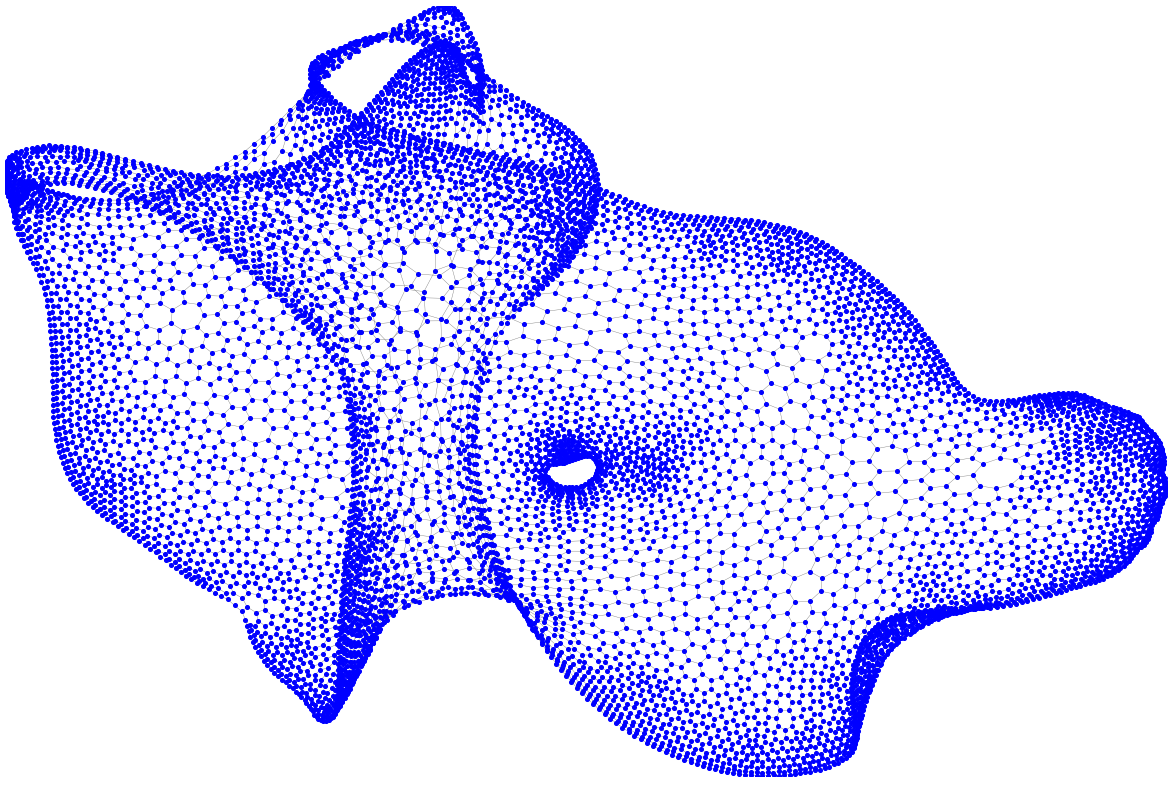} & \includegraphics[width=0.16\linewidth]{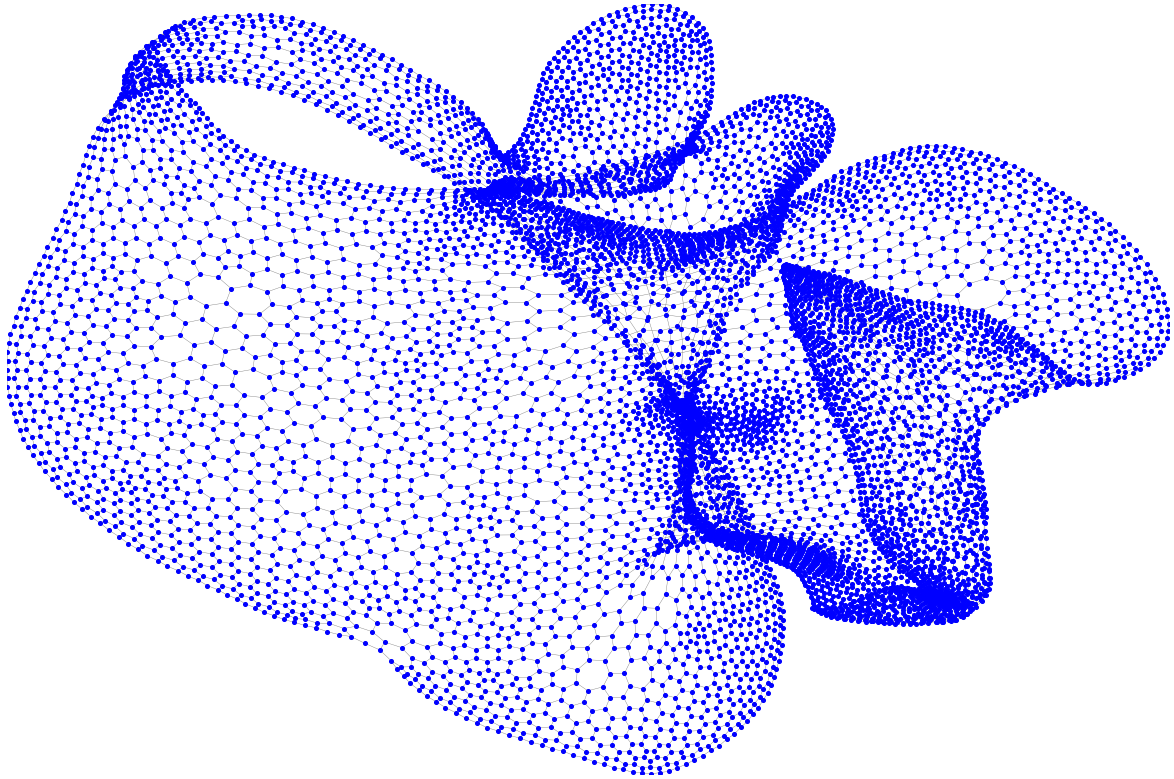} & \includegraphics[width=0.16\linewidth]{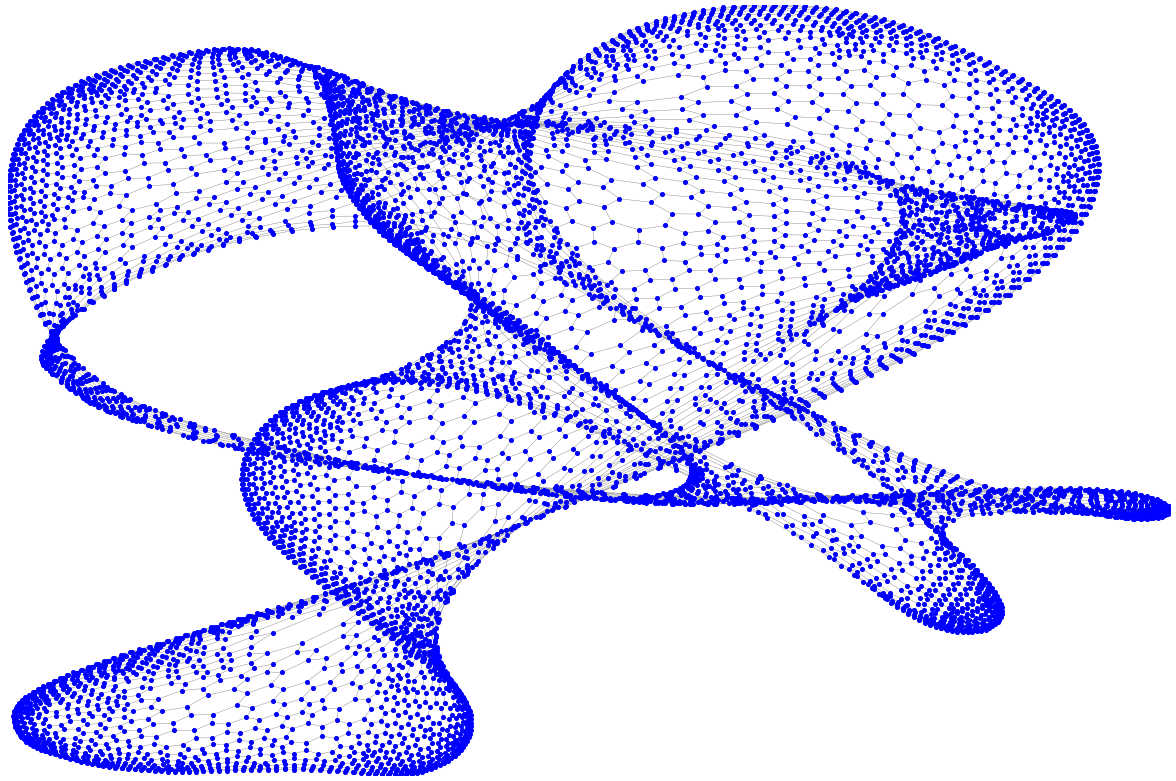} & \includegraphics[width=0.16\linewidth]{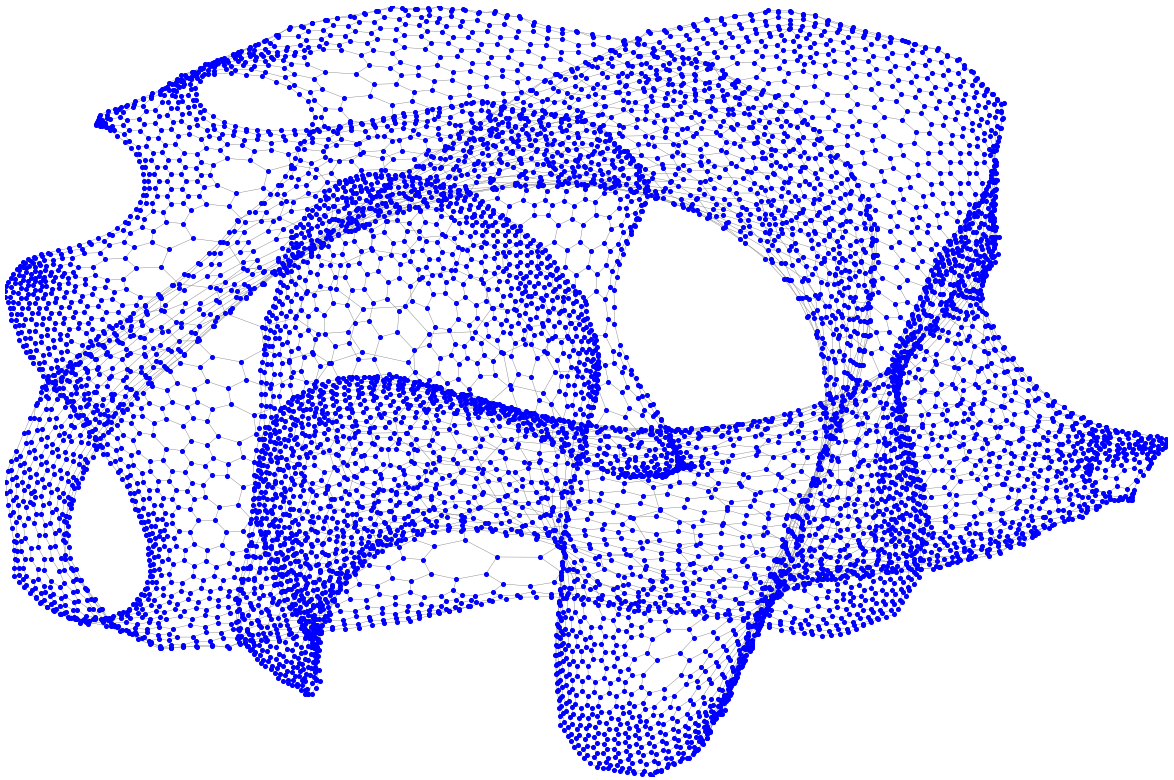} &
         \includegraphics[width=0.16\linewidth]{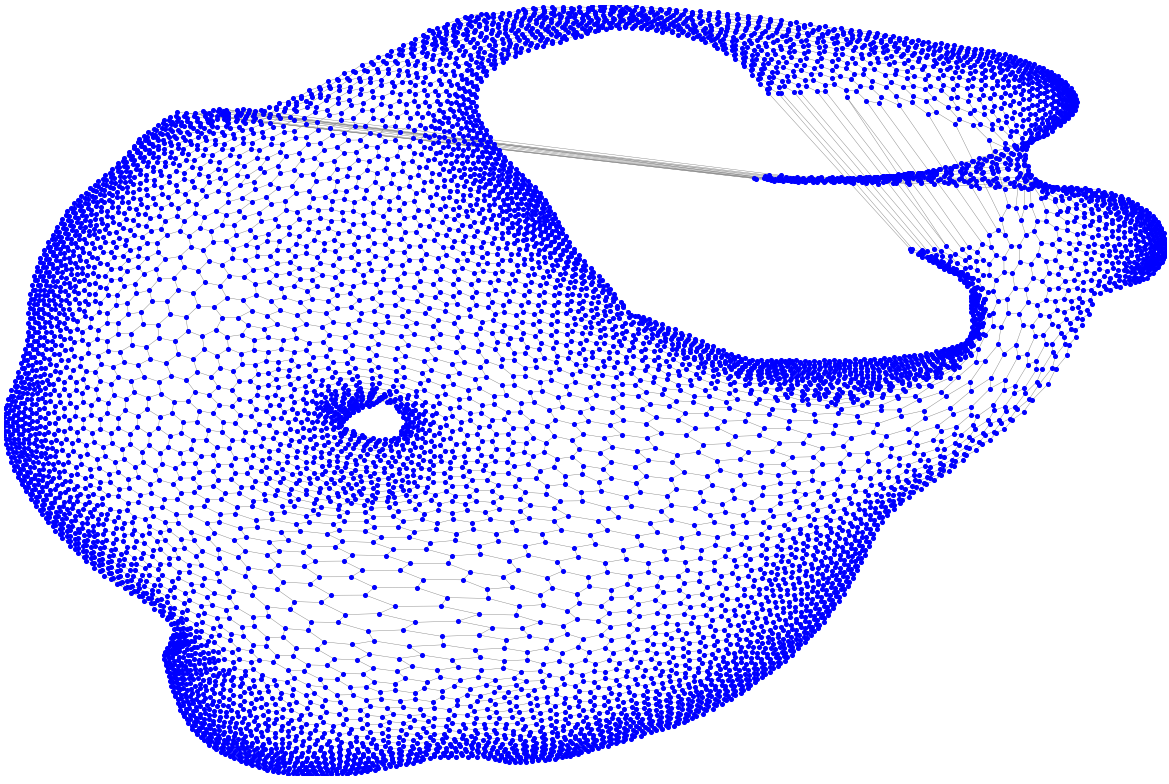} \\
         \midrule
         \rotatebox{90}{tube2} &  \includegraphics[width=0.16\linewidth]{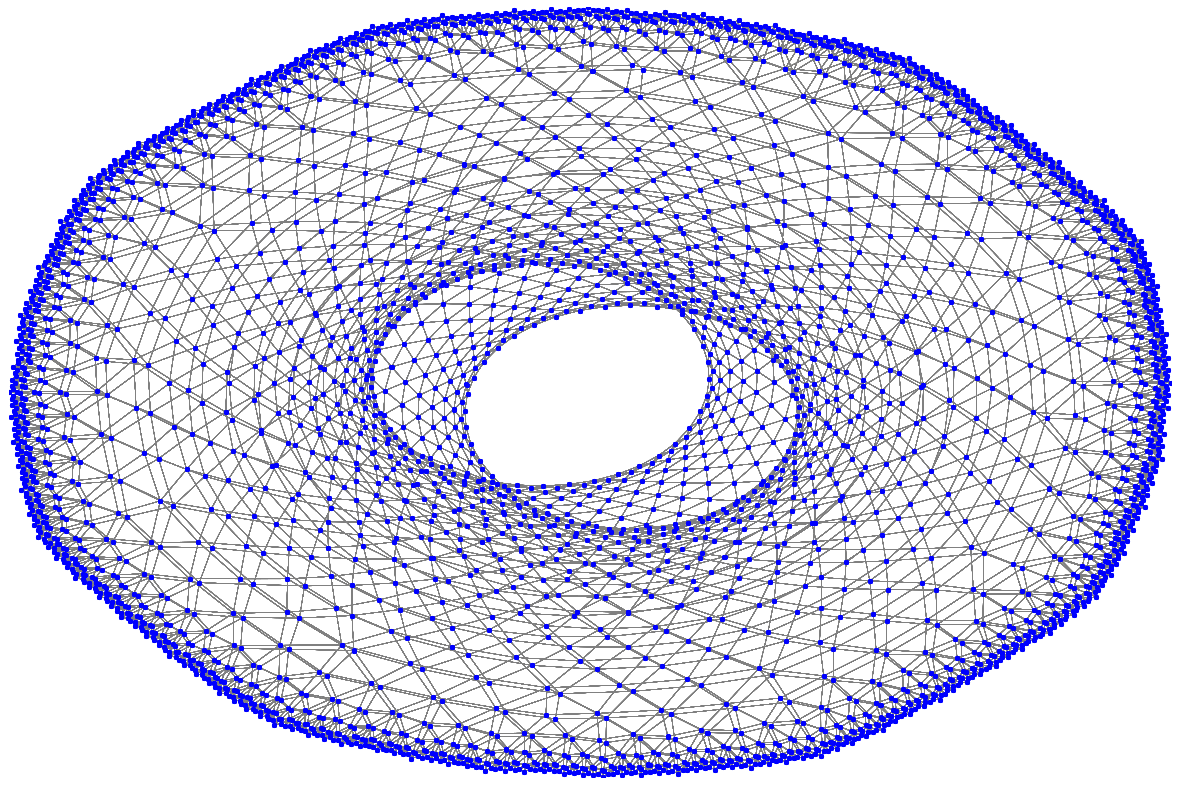} & \includegraphics[width=0.16\linewidth]{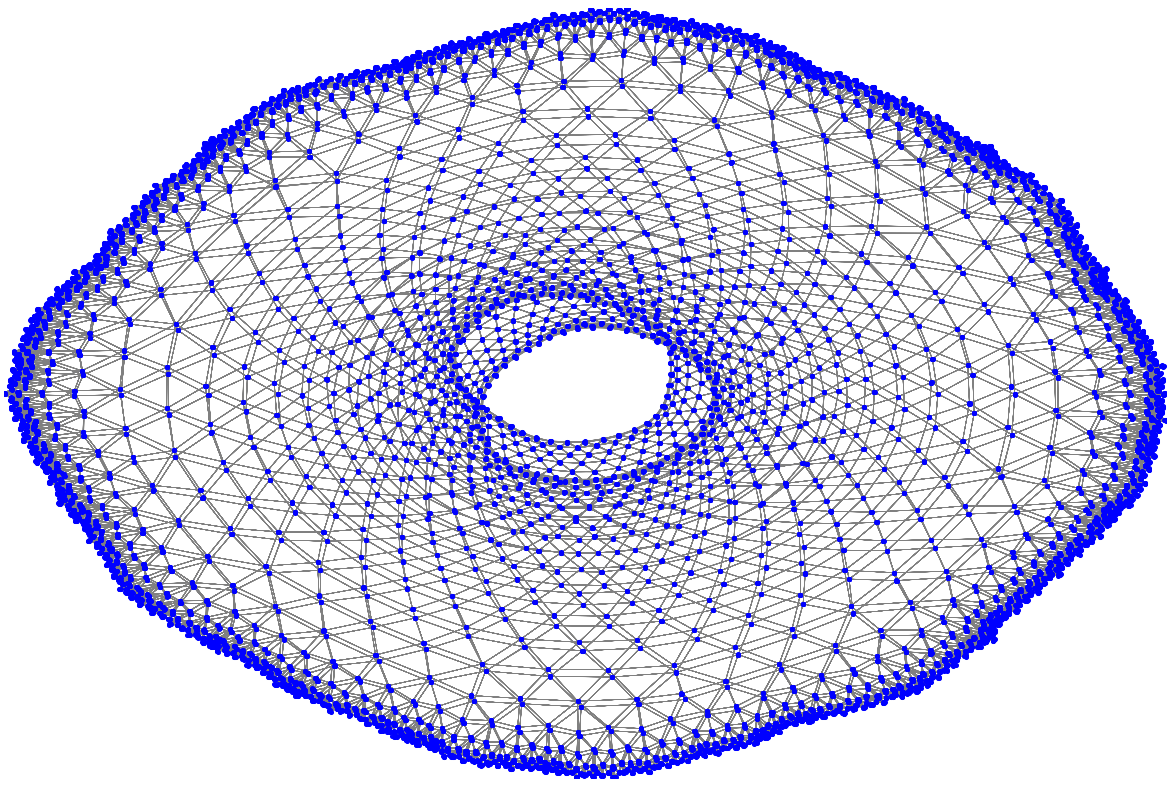} & \includegraphics[width=0.16\linewidth]{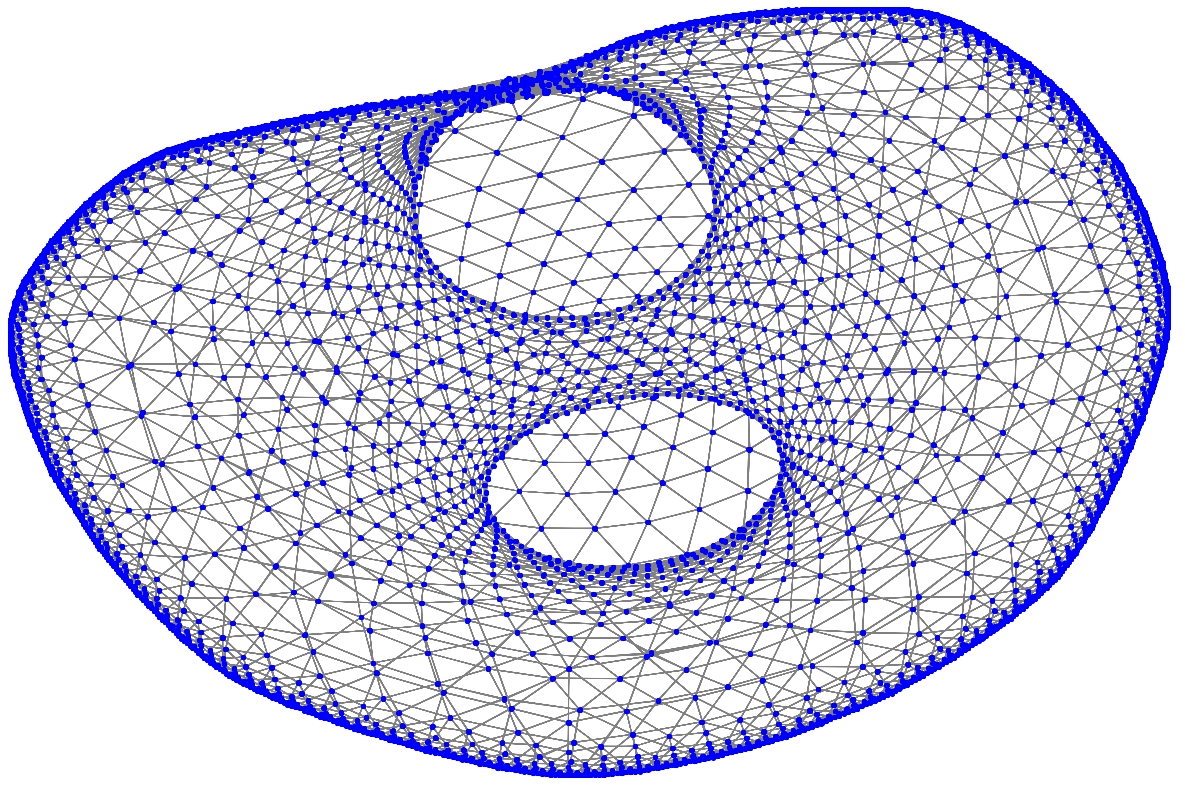} & \includegraphics[width=0.16\linewidth]{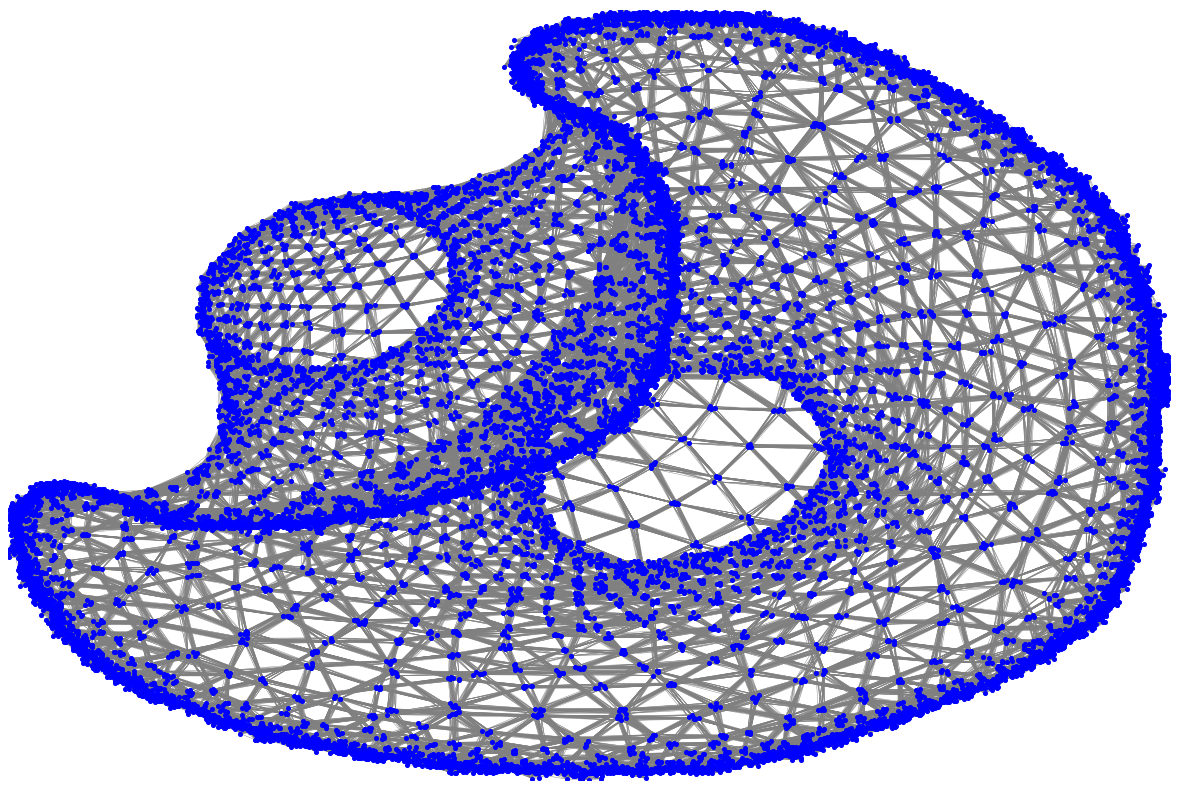} &
         \includegraphics[width=0.16\linewidth]{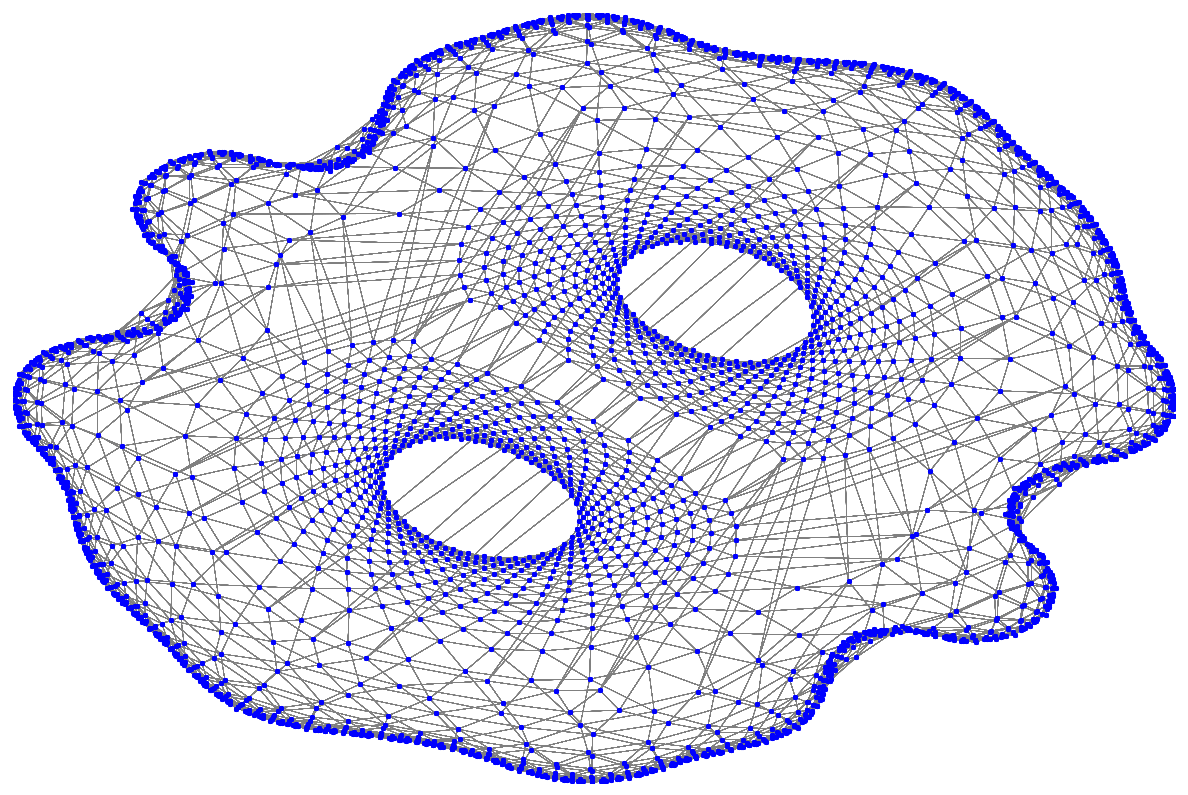} \\
         \midrule
         \rotatebox{90}{finance256} &  \includegraphics[width=0.16\linewidth]{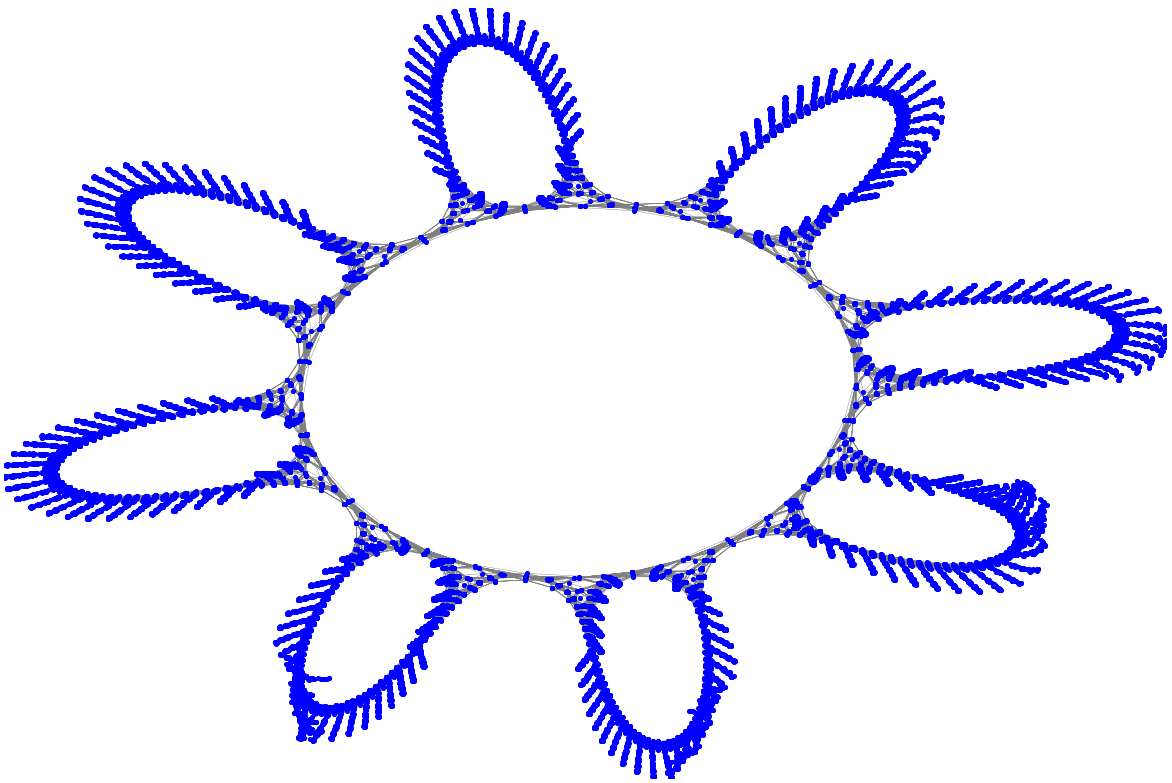} & \includegraphics[width=0.16\linewidth]{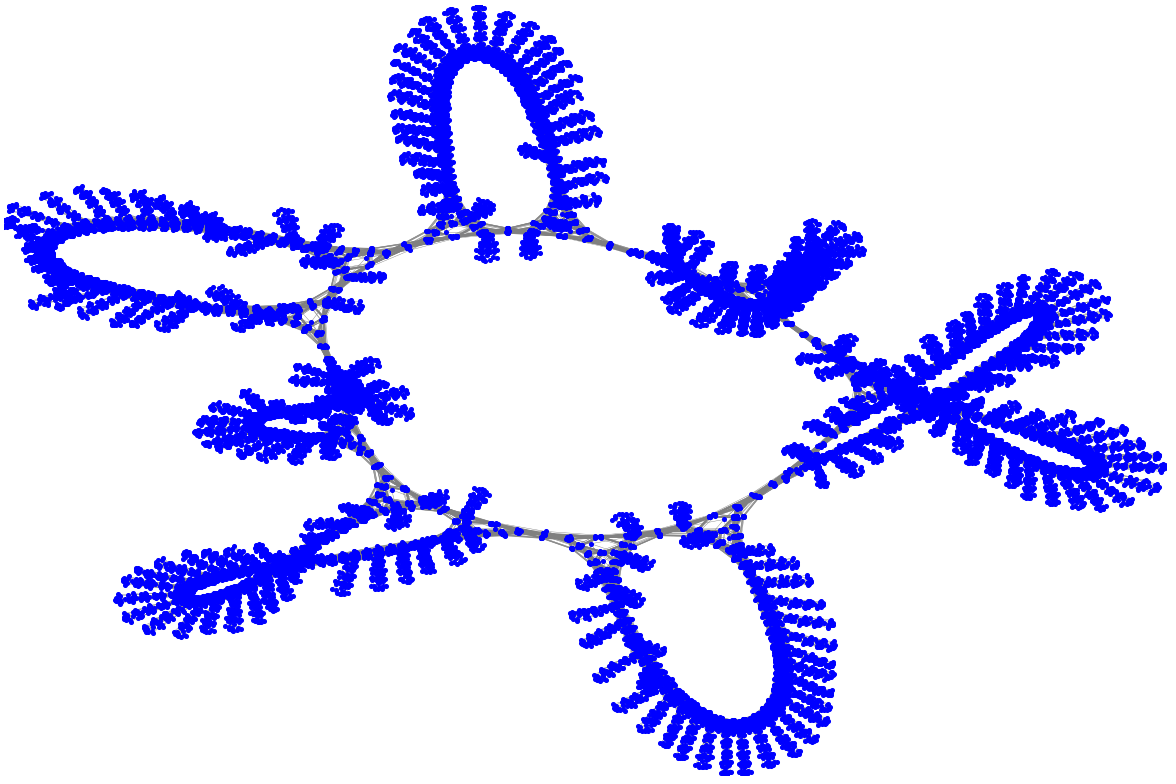} & \includegraphics[width=0.16\linewidth]{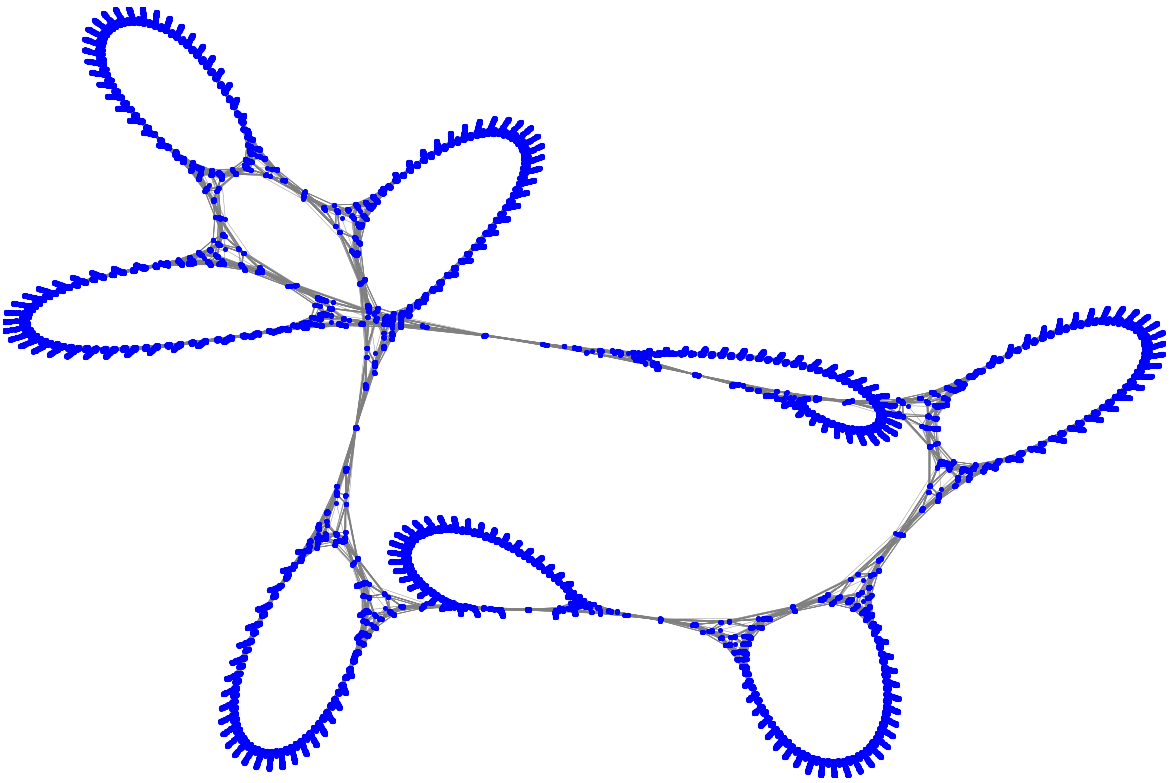} & \includegraphics[width=0.16\linewidth]{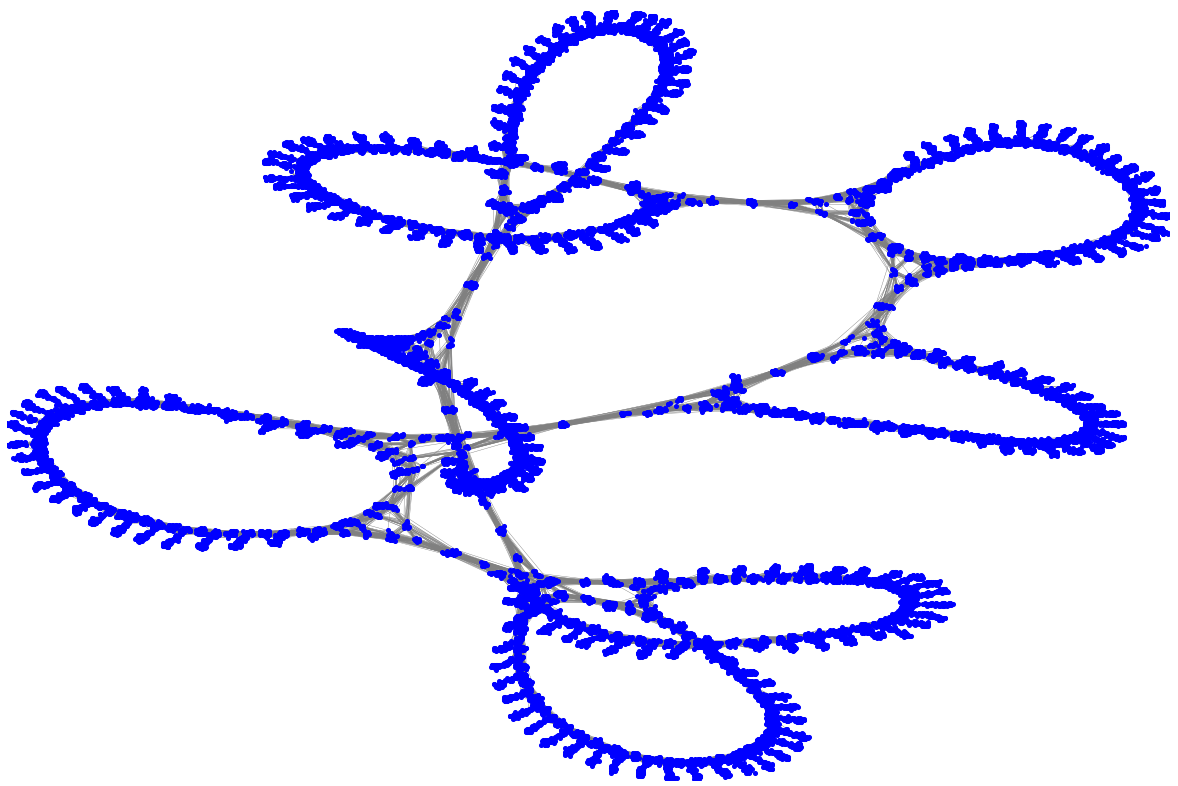} &
         \includegraphics[width=0.16\linewidth]{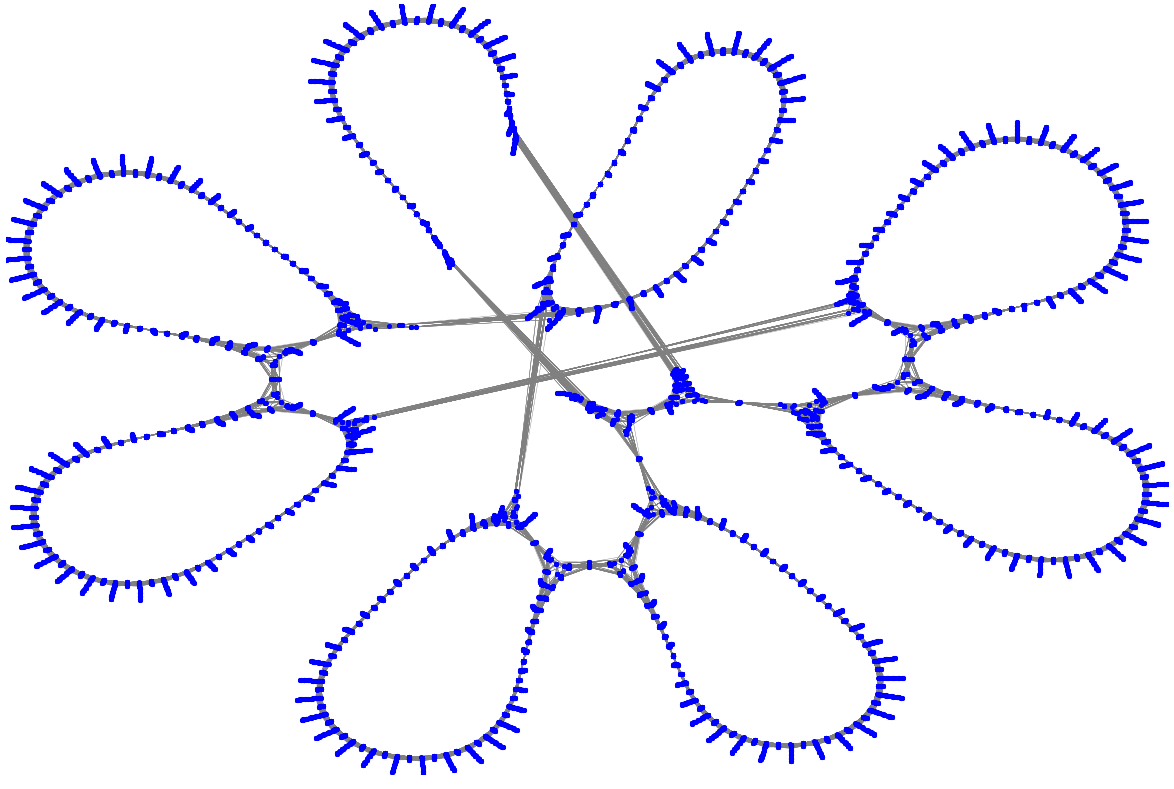} \\
         \bottomrule
    \end{tabular}
    \label{tab:layouts}
\end{table}

\subsubsection{Qualitative Comparison}
We show the quality of the generated layouts by different tools in Table \ref{tab:layouts}. We observe that the layouts generated by tsNET is the most readable than others. The quality of layouts generated by BatchLayout are close to tsNET. Note that BatchLayout generates these layouts within a very short time compared to tsNET. The layouts of other methods are also comparable. Different people may have different perception for visual quality. Thus, in the next section, we also analyze the generated layouts using quantitative measures.

\begin{table}[!htb]
\centering
\caption{Edge crossing and edge uniformity measures of different graph visualization methods for two representative graphs shown in Table \ref{tab:layouts}. A lower value is better for both edge crossing and edge uniformity measures. BL - BatchLayout, BLBH - Barnes-Hut based BatchLayoout, FA2BH - Barnes-Hut based ForceAtlas2, OO - OpenOrd.}
\arrayrulecolor{black}
\begin{tabular}{!{\color{black}\vrule}c!{\color{black}\vrule}c!{\color{black}\vrule}c!{\color{black}\vrule}c!{\color{black}\vrule}c!{\color{black}\vrule}c!{\color{black}\vrule}c!{\color{black}\vrule}c!{\color{black}\vrule}c!{\color{black}\vrule}c!{\color{black}\vrule}c!{\color{black}\vrule}} 
\arrayrulecolor{black}\cline{1-1}\arrayrulecolor{black}\cline{2-11}
\multirow{2}{*}{Graphs} & \multicolumn{5}{c!{\color{black}\vrule}}{Edge Crossing$\downarrow$} & \multicolumn{5}{c!{\color{black}\vrule}}{Edge Uniformity$\downarrow$}  \\ 
\cline{2-11}
                        & BL       & BLBH     & FA2BH    & OO       & tsNET       & BL   & BLBH & FA2  & OO   & tsNET                          \\ 
\arrayrulecolor{black}\cline{1-1}\arrayrulecolor{black}\cline{2-11}
grid2\_dual             & 1,043 & 515  & 596   & 2,549 & 0        & 0.40 & 0.35 & 0.58 & 0.55 & 0.26                           \\ 
\hline
3elt dual               & 2,400 & 3,851 & 5,385 & 7,354 & 692      & 0.41 & 0.42 & 0.64 & 0.53 & 0.62                           \\
\hline
\end{tabular}
\label{tab:ecandeu}
\arrayrulecolor{black}
\end{table}

\begin{table}[!htb]
\centering
\caption{Stress and Neighborhood preservation measures of different graph visualization methods for two representative graphs shown in Table \ref{tab:layouts}. A lower value is better for stress measure whereas a higher value is better for Neighborhood preservation measure. BL - BatchLayout, BLBH - Barnes-Hut based BatchLayoout, FA2BH - Barnes-Hut based ForceAtlas2, OO - OpenOrd.}
\arrayrulecolor{black}
\begin{tabular}{!{\color{black}\vrule}c!{\color{black}\vrule}c!{\color{black}\vrule}c!{\color{black}\vrule}c!{\color{black}\vrule}c!{\color{black}\vrule}c!{\color{black}\vrule}c!{\color{black}\vrule}c!{\color{black}\vrule}c!{\color{black}\vrule}c!{\color{black}\vrule}c!{\color{black}\vrule}} 
\arrayrulecolor{black}\cline{1-1}\arrayrulecolor{black}\cline{2-11}
\multirow{2}{*}{Graphs} & \multicolumn{5}{c!{\color{black}\vrule}}{Stress$\downarrow$} & \multicolumn{5}{c!{\color{black}\vrule}}{Neighborhood Preservation$\uparrow$}  \\ 
\cline{2-11}
                        & BL     & BLBH   & FA2    & OO     & tsNET        & BL   & BLBH & FA2  & OO   & tsNET                                    \\ 
\arrayrulecolor{black}\cline{1-1}\arrayrulecolor{black}\cline{2-11}
grid2\_dual             & 4.1E+5 & 2.4E+5 & 4.3E+5 & 4.5E+5 & 1.4E+5       & 0.50 & 0.60 & 0.53 & 0.43 & 0.73                                     \\ 
\hline
3elt dual               & 3.7E+6 & 6.1E+6 & 6.6E+6 & 7.5E+6 & 3.1E+6       & 0.55 & 0.51 & 0.37 & 0.36 & 0.64                                     \\
\hline
\end{tabular}
\label{tab:sandnp}
\arrayrulecolor{black}
\end{table}
\subsubsection{Quantitative Comparison}
To assess the quantitative measure of visualization shown in Table \ref{tab:layouts}, we choose grid2\_dual and 3elt\_dual graphs. We use four graph visualization measures \citep{de2019multi,rahman2020batchlayout}, namely, (i) the number of edge crossing, (ii) edge length uniformity, (iii) stress, and (iv) neighborhood preservation. A lower value means a better results for edge crossing, edge length uniformity and stress measures. On the other hand, a higher value means a better result for neighborhood preservation measure. All these measures are widely used in the literature to assess the layouts quantitatively. We report the results in Tables \ref{tab:ecandeu} and \ref{tab:sandnp}. We observe that tsNET wins in edge crossing, stress and neighborhood measures. One version of BatchLayout always shows runnerup results. For edge uniformity measure, BatchLayout shows competitive results. BatchLayout shows this competitive performance with a very low runtime compared to tsNET.

\section{Discussions}
\quad Taking into account all of the data and analysis, we can see that all of the algorithms tested have certain strengths and weaknesses. With this, we can find a use case for all of the algorithms. 

\textbf{High-dimensional Vector Data.} With the focus on solving the crowding-out problem, t-SNE is the best choice for reducing data-sets where a more uniformly distributed embedding is desired. Furthermore, with the most stable memory usage among algorithms, it is well applied to use cases where predictable memory usage is favorable. However, we recommend that the original serial version of t-SNE is not used, as it fails to scale to large data-sets, with its high memory consumption and high computational complexity. Therefore it is recommended that if it is used, that one of its more optimized versions be used. Being based upon t-SNE with the same KNN tree construction, LargeVis shares with it certain characteristics, such as it shows more stable memory usage. However, LargeVis is significantly less computationally complex than t-SNE, allowing it to scale to data-sets with greater data points. With the focus of LargeVis being to improve runtime as compared to t-SNE, it certainly delivers. Therefore, we believe that LargeVis should be used for applications where stable memory usage, high scalablity, and the preservation of the data's local structure are desired. Moving on to UMAP, we are given faster runtime than both t-SNE and LargeVis, but at the cost of less stable memory usage. Providing similar embedding quality as compare to LargeVis, UMAP should be used as an alternative for LargeVis use cases when the stability of memory is not a factor, as it provides similar results in nearly half the time. Finally, we are left with the fastest and most memory efficient of all algorithms tested, that being TriMap. With its significantly lower runtimes and memory usage TriMap certainly outclasses all other tested algorithms in all tests. However, TriMap is somewhat limited by its design specifications, as it is primarily geared towards maintaining the global structure of data, rather than local structure. With relatively unstable memory usage, the predictability of TriMap is less than that of algorithms like LargeVis and t-SNE. Therefore, we believe that the best use cases for TriMap are whenever maintaining the global structure of data is the goal or the size of the data-set is too great for any other algorithm to realistically process.

\textbf{Graph Data.} For graph data visualization, tsNET methods take a long time to generate a layout of a medium-scale graph and it also consumes a high amount of memory. Nevertheless, the layouts generated by tsNET are of good quality. If users have enough time and memory, then tsNET can be used to generate a good quality layout. OpenOrd is a high-performance graph visualization tool that can be run on a distributed system to generate the layout of a big graph. Thus, when we need to generate graph layout quickly, we can use OpenOrd though the quality may not be good. ForceAtlas2 is integrated with Gephi software which is user-friendly and scientists/researchers from other research domains can also use ForceAtlas2 without coding skill. Due to maintaining the structure of the Gephi framework, this method consumes a high amount of memory which reduces its usability on a large graph. Finally, the BatchLayout method generates good quality layout very quickly compared to other methods. It can also generate layouts within a reasonable time for graphs having millions of vertices. Thus, users can also consider using BatchLayout which simultaneously produces good quality layouts within a short time consuming less memory.

\section{Conclusion and Future Work}
Data visualization provides an important insight about the high-dimensional data in low-dimensional space. This is a precursor for data-mining and information retrieval. In this survey paper, we analyzed several state-of-the-art visualization techniques. To compare the results, we mainly focused on runtime, memory consumption and visual quality using benchmark datasets. From our analysis, we can recommend users to chose a visualization method for a given set of computing resources.

With the breadth and extensibility of established algorithms such as t-SNE and UMAP, the results for outputs can vary greatly based on the tuning modifications, making it a component worth investigating further. The quality of graph layouts can also vary based on the selection of different hyper-parameters where users should be careful and follow the guidelines of parameter sensitivity in the papers. Finally, an extension of our analysis to include modified implementations of algorithms already discussed, such as those of t-SNE and BatchLayout, may provide more depth to the strength of using more established algorithms.

\bibliographystyle{spbasic}
\bibliography{main}
\end{document}